%% file: root.tex
\documentclass[10pt,journa,doublecolumn]{IEEEtran} 
\IEEEoverridecommandlockouts

\usepackage{amsfonts, amsmath,amssymb,amsthm}
\usepackage{algorithm, algorithmic}
\usepackage{array}
\usepackage{cases}
\usepackage{changepage}
\usepackage[compress]{cite}
\usepackage{epstopdf}
\usepackage[english]{babel}
\usepackage{etoolbox}
\usepackage{graphicx}
\usepackage{multirow}
\usepackage{physics} 	
\usepackage{subfigure}
\usepackage{setspace}
\usepackage{url}  
\usepackage[hyphenbreaks]{breakurl} 
\usepackage{lipsum}
\usepackage{nicefrac}	
\usepackage[table,xcdraw]{xcolor}	
\usepackage{textcomp}
\usepackage{makecell}
\usepackage[font=small]{caption} 
\usepackage{booktabs} 
\usepackage{mathrsfs}
\usepackage{indentfirst} 
\usepackage{textcomp} 
\usepackage{pifont}		
\usepackage{threeparttable}
\usepackage[right]{lineno}  

\def\ps@headings{%
\def\@oddhead{\mbox{}\scriptsize\rightmark \hfil \thepage}%
\def\@evenhead{\scriptsize\thepage \hfil \leftmark\mbox{}}%
\def\@oddfoot{}%
\def\@evenfoot{}}
\makeatother
\makeatletter
\patchcmd{\@thm}
  {\trivlist}
  {\list{}{\leftmargin=\thm@leftmargin\rightmargin=\thm@rightmargin}}
  {}{}
\patchcmd{\@endtheorem}
  {\endtrivlist}
  {\endlist}
  {}{}
\newlength{\thm@leftmargin}
\newlength{\thm@rightmargin}

\newcommand{\xnewtheorem}[3]{%
  \newenvironment{#3}
    {\thm@leftmargin=#1\relax\thm@rightmargin=#2\relax\begin{#3INNER}}
    {\end{#3INNER}}%
  \newtheorem{#3INNER}%
}

\makeatother

\newtheoremstyle{indentedupright}{3pt}{3pt}{} {}{\bfseries}{.}{.5em}{} 
\newtheoremstyle{indenteditalic}{3pt}{3pt}{\itshape} {}{\bfseries}{.}{.5em}{} 

\theoremstyle{indenteditalic}
\xnewtheorem{0pt}{0pt}{oldpro}{Conventional Property}
\xnewtheorem{0pt}{0pt}{newpro}{New Property}
\xnewtheorem{0pt}{0pt}{obser}{Observation}
\xnewtheorem{0pt}{0pt}{pro}{Property}
\xnewtheorem{0pt}{0pt}{lem}{Lemma}
\xnewtheorem{0pt}{0pt}{corl}{Corollary}
\xnewtheorem{0pt}{0pt}{addcon}{Additional Constraint}

\newcommand{\romu}[1]{\uppercase\expandafter{\romannumeral #1\relax}} 
\newcommand{\roml}[1]{\lowercase\expandafter{\romannumeral #1\relax}}    

\renewcommand{\baselinestretch}{1}

\begin{document}
\title{\LARGE ML-ARIS: Multilayer Underwater Acoustic Reconfigurable Intelligent Surface with High-Resolution Reflection Control}
\author{\IEEEauthorblockN{Lina Pu\IEEEauthorrefmark{1}, Yu Luo\IEEEauthorrefmark{2}, Aijun Song\IEEEauthorrefmark{3}}\\
\IEEEauthorblockA{\IEEEauthorrefmark{1}Department of Computer Science, University of Alabama, Tuscaloosa, AL 35487\\}
\IEEEauthorrefmark{2}Department of Electrical and Computer Engineering, Mississippi State University, Mississippi State, MS, 39759\\
\IEEEauthorrefmark{3}Department of Electrical and Computer Engineering, University of Alabama, Tuscaloosa, AL 35487\\
Email: lina.pu@ua.edu, yu.luo@ece.msstate.edu, song@eng.ua.edu}

\maketitle
\begin{abstract}
\label{:Abstract}
This article introduces a multilayered acoustic reconfigurable intelligent surface (ML-ARIS) architecture designed for the next generation of underwater communications. ML-ARIS incorporates multiple layers of piezoelectric material in each acoustic reflector, with the load impedance of each layer independently adjustable via a control circuit. This design increases the flexibility in generating reflected signals with desired amplitudes and orthogonal phases, enabling passive synthetic reflection using a single acoustic reflector. Such a feature enables precise beam steering, enhancing sound levels in targeted directions while minimizing interference in surrounding environments. Extensive simulations and tank experiments were conducted to verify the feasibility of ML-ARIS. The experimental results indicate that implementing synthetic reflection with a multilayer structure is indeed practical in real-world scenarios, making it possible to use a single reflection unit to generate reflected waves with high-resolution amplitudes and phases.
\end{abstract}



\section{Introduction}
\label{sec:Intro}
Recent advancements in reconfigurable intelligent surfaces (RIS) have attracted significant attention for their potential to enhance the performance of radio communication systems~\cite{ahmed2024active, yang2023beyond}. These surfaces utilize a large array of elements, which can be electronically adjusted to actively control the phase, amplitude, and polarization of incoming signals. This capability allows for precise steering of reflected waves toward designated receivers while minimizing interference in specified zones. As a result, RF-RIS promises to boost signal strength, expand coverage, and streamline interference management in the next generation of wireless networks~\cite{wu2024intelligent}.

Compared with active acoustic relays or repeaters, passive operation of RIS offer a fundamentally different operational trade-off: they shape the acoustic propagation environment without generating new acoustic signals. This makes RIS particularly attractive in scenarios where power availability, acoustic self-noise, or system visibility are critical constraints. Representative applications include long-duration seabed sensor networks and observatories, where battery replacement is costly~\cite{sutton2005ocean}; infrastructure-assisted coverage extension for autonomous underwater vehicle (AUV) operations, where passive surfaces can redirect energy into desired regions; and covert or environmentally sensitive deployments, where minimizing acoustic emissions is desirable~\cite{luo2014challenges}. In these settings, RIS complements active systems by providing a low-power, low-complexity means of enhancing link reliability and spatial selectivity, rather than replacing transmitters or receivers. This motivates the need for advanced reflection control and environmental adaptability.

Existing RIS research predominantly focuses on radio frequency (RF) communications, which face significant range limitations in aquatic settings due to substantial absorption losses in water~\cite{shaw2006experimental}. As a result, acoustic communication has been the principal method for mid-range and long-range underwater communication. Due to the fundamental differences between RF and acoustic signals, current RF-RIS designs are unsuitable for direct application in underwater acoustic RIS (UA-RIS). Currently, only a few studies explored the design of UA-RIS~\cite{sun2022high, wang2023designing, luo2025underwater}, highlighting a significant opportunity for advancements and breakthroughs in the field.

Our field experiments have confirmed the potential of UA-RIS to enhance the range and data rate of underwater acoustic communications~\cite{luo2024experimental}. Nevertheless, transitioning UA-RIS from theory concepts to practical application still needs to overcome a series of unique challenges:
\vspace{0.1cm}
\begin{adjustwidth}{-0.71cm}{0cm}
\begin{description}
\setlength{\labelsep}{-0.95em}
\itemsep 0.07cm
  \item[a)] \textbf{Varying impedance of acoustic reflector:} The impedance of piezoelectric materials like lead zirconate titanate (PZT) varies with frequency, water depth, temperature, and acoustic wave angle, unlike RF reflectors, which depend only on frequency. Therefore, the matching circuit for acoustic reflectors must be electronically reconfigurable to adapt to these environmental changes.
  \item[b)] \textbf{Low frequency of acoustic signal:} Acoustic signals in underwater communication are typically below 100\,kHz, challenging traditional phase-shifting methods that use varactors, which don't provide enough capacitance variation. This necessitates developing new phase-shifting mechanisms for UA-RIS to effectively control wave reflection.
  \item[c)] \textbf{Size and weight constraint:} Unlike RF-RIS, which can accommodate thousands of reflection unit cells on a printed circuit board (PCB), acoustic reflectors need to be larger and heavier to resonate with mechanical waves at low frequencies. This constraint significantly reduces the number of reflecting units in UA-RIS, requiring innovative designs to maximize directional gain with fewer units.
\end{description}
\end{adjustwidth}
\vspace{0.1cm}

To address these challenges, we introduce a multilayered acoustic reconfigurable intelligent surface (ML-ARIS) featuring a stacked architecture to enable flexible acoustic reflection. In ML-ARIS, the reflector comprises multiple stacked PZT disks, each with independently adjustable load impedance via a control circuit. This design increases the flexibility in generating reflected signals with desired amplitudes and orthogonal phases. By programming the control circuit, the superposed wave from each reflector can achieve a wide and continuously tunable range of phase and amplitude values, subject to the physical and circuit constraints of the multilayer reflector. Consequently, ML-ARIS can execute passive synthetic reflection on a single acoustic reflector, enabling precise beam steering and significantly boosting wave intensity toward targeted areas.

Unlike multi-bit coding RF-RIS, which typically produces a very limited number of phase shift states and lacks control over the strength of the reflected signal~\cite{kim2023rotated, zhang2019dynamically}, ML-ARIS with synthetic reflection can generate a high-resolution reflection coefficient. Consequently, ML-ARIS is capable of performing advanced array processing methods such as robust capon beamforming (RCB) and minimum variance distortionless response (MVDR). These techniques enhance the intensity of reflected waves in targeted directions while mitigating unintended interference in surrounding environments.

In addition to the multilayer structure, an enhanced matching network is implemented to address the varying impedance of piezoelectric materials in dynamic environments. It utilizes a three-tier subnetwork architecture, each tier comprising a high-pass L-type matching circuit. A microcontroller (MCU) coordinates the interconnections between tiers, adapting to changes in the reflector's impedance due to temperature fluctuations and the incident angle of acoustic waves. For optimal performance, the tiers operate in a cascaded manner rather than independently. The parameters of all components in the matching circuit, including capacitance and inductance, are determined by solving an optimization problem.

We conducted extensive experiments in a tank environment, as well as simulations using the COMSOL platform, to evaluate the performance of ML-ARIS at around 28\,kHz and 41\,kHz. The tests demonstrated that with the proposed matching circuit, the magnitude of the reflection coefficient at the target load remains below 0.15 as the water temperature ranges from 9$^\circ$C to 22$^\circ$C and the acoustic signal frequency varies between 27.5\,kHz and 28.5\,kHz at different water depths. Both the simulation and tank experiment results confirm that the stacked architecture with synthetic reflection effectively manipulates the amplitude and phase of reflected waves on a single reflection unit. Compared to multi-bit coding schemes, ML-ARIS with synthetic reflection generates significantly lower side lobes in the reflected beam, enabling more devices to share the acoustic channel without causing interference. In particular, our work makes three key contributions:
\vspace{0.1cm}
\begin{adjustwidth}{-0.71cm}{0cm}
\begin{description}
\setlength{\labelsep}{-0.95em}
\itemsep 0.07cm
	\item[a)] A multilayer structure-based ARIS is proposed for underwater acoustic communications. ML-ARIS enables synthetic reflection on each reflector, allowing for creation of reflected signal with tailored amplitude and phase, enabling precise beam steering through advanced array processing. 
    \item[b)] We introduce a specialized matching network to accommodate variations in the reflector's impedance caused by environmental changes. This ensures consistent impedance matching along the transmission line, crucial for accurate synthetic reflection.
     \item[c)] Both tank experiments and COMSOL-based simulations are conducted to assess the feasibility of ML-ARIS. The results validate that the stacked architecture effectively controls the reflected wave, significantly boosting signal strength in targeted directions and minimizing interference.
\end{description}
\end{adjustwidth}
\vspace{0.1cm}

The remainder of this paper is organized as follows: Section~\!\ref{sec:Related} presents the related work on RIS. Section~\!\ref{sec:Moti} motivates the structure of multilayered UA-RIS. The hardware design of ML-ARIS is detailed in Section~\!\ref{sec:Hardware}. Section~\!\ref{sec:MatCir} addresses strategies for enhancing the performance of the matching circuit. Finally, in Section~\!\ref{sec:PerEva}, we evaluate the performance of the proposed UA-RIS through simulation and experimental validation. Section~\!\ref{sec:Con} concludes our work.


\section{Related Work}
\label{sec:Related}
In recent years, extensive research has been undertaken to optimize RF-RIS performance, address implementation challenges, and explore its applicability across diverse communication scenarios. For instance, in \cite{tang2020wireless}, three distinct free-space path loss models were developed to analyze the relationship between the path loss in RIS-assisted wireless communications and variables such as the distances, RIS size, and radiation patterns of reflected waves. These models were validated through experiments with fabricated RIS, paving the way for future theoretical insights and practical innovations in the field.

The work presented in \cite{huang2019reconfigurable} investigated the energy efficiency of RIS-assisted communication systems. By optimizing both the transmitter power and the phase shifts of RIS elements, their theoretical analysis suggests that an RIS-assisted network could achieve up to 300\% higher energy efficiency compared to conventional multi-antenna amplify-and-forward relaying systems, without compromising quality of service. Additionally, \cite{ren2023deployment} discussed deployment strategies for RIS in millimeter-wave (mmWave) multi-cell networks. This study introduced environment-aware algorithms to optimize RIS placement, enhancing coverage and capacity while mitigating the signal blockages typical in mmWave communications.

In \cite{jian2022reconfigurable}, a survey of RF-RIS hardware development is provided, highlighting challenges and innovations in designing cost-effective RIS elements. The role of metametamaterials and programmable metasurfaces in achieving dynamic control over EM waves was emphasized. Techniques for achieving flexible phase shifts using varactor diodes were examined in \cite{boccia2002application}, \cite{tan2016increasing}, and \cite{pei2021ris}. These works involve placing a varactor diode at the radiating edge of a patch antenna and adjusting its reverse voltage to electronically control the RIS's phase response, allowing for a broad tunable phase control.

Currently, only a limited number of studies have investigated the application of RIS techniques in underwater environments. For instance, \cite{eid2023enabling} demonstrated the feasibility of using backscattered signals for long-distance passive communications with a Van Atta array. By switching the load of each reflector between short and open circuit states, the polarization of the reflected wave is toggled. Consequently, the strength of the received signal, which is the superposition of emitted and reflected waves, exhibits variations. Sea experiments indicate that the Van Atta backscatter array can achieve a round-trip communication range exceeding 300 meters with a bit error rate (BER) of $10^{-3}$.

In \cite{sun2022high}, the enhancement of communication performance using UA-RIS was examined. In this design, each reflection unit incorporates a piezoelectric coil positioned between two metal plates, with the resonant frequency adjusted by the reverse voltage applied to a varactor. Meanwhile, \cite{wang2023designing} explored the design of UA-RIS capable of responding to a wide range of frequencies. This work addressed  element dispersion within each reflector and array dispersion throughout the acoustic RIS. By counteracting these dispersions, the directivity of the reflected waves becomes frequency-independent, allowing wideband signals to achieve uniform beam patterns across various frequency components. The efficacy of this architecture was validated through Bellhop and COMSOL simulators.


\section{Motivation of Multilayered UA-RIS}
\label{sec:Moti}
This section outlines the motivation behind adopting a multilayered UA-RIS. We begin by discussing the importance of simultaneously manipulating both the amplitude and phase of reflected waves in UA-RIS applications. Afterward, the inherent challenges in achieving flexible reflection control are clarified. Finally, we briefly review our previous solution for implementing such control and highlight its limitations.

\subsection{Necessity of Flexible Reflection Control}
The underwater acoustic channel is a crucial resource, vital not only for manmade facilities but also for marine animals that rely on acoustic signals for hunting, communication, and navigation~\cite{luo2014challenges}. However, the absorption attenuation of sound signals increases significantly with frequency, limiting the available bandwidth for mid-range and long-range acoustic applications to just tens of kilohertz. As the number of marine facilities grows, the need to efficiently utilize the acoustic channel without adversely impacting the surrounding marine life becomes a critical challenge for future acoustic systems. 

\begin{figure}[htb]
\centerline{\includegraphics[width=8.0cm]{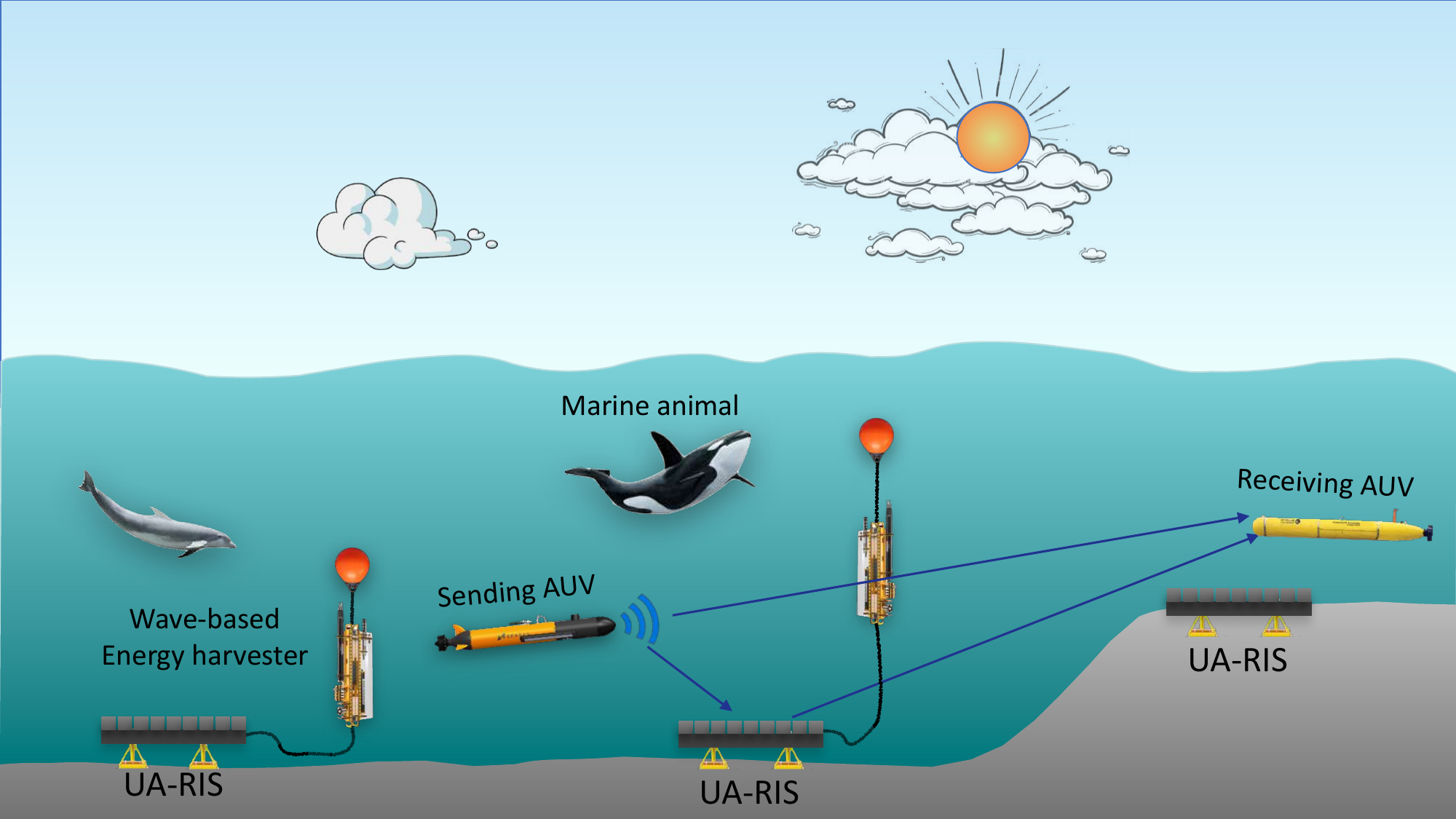}}
  \caption{Scenario of UA-RIS assisted underwater acoustic systems for eco-friendly communications.}\label{fig:Scenario}
\end{figure}

To meet the above objective, multiple UA-RIS facilities can be deployed on the seafloor at predetermined locations. As illustrated in Fig.~\!\ref{fig:Scenario}, each facility is connected to a wirewalker~\cite{del2021wirewalker}, a fully passive device that harvests energy from ocean-wave motion.  With appropriate system design, UA-RIS units can steer the source signal toward the intended receiver and coherently combine the reflected acoustic waves, thereby significantly enhancing the signal quality. In addition, every UA-RIS can naturally work as a passive sonar system. As a result, the relative directions of the sender-reflector and receiver-reflector paths can be estimated using angle-of-arrival (AoA) measurements.

The goal of UA-RIS is to introduce a controllable, phase-adjustable propagation component whose amplitude and delay are engineered to (a) combine constructively with the existing arrivals at the receiver, and (b) provide an additional high-SNR path that the receiver can either resolve or coherently combine to enhance overall communication performance. This capability will allow the source level of acoustic transmissions to be reduced without degrading communication performance, ultimately supporting more environmentally sustainable underwater communication.

To achieve optimal performance, RIS-assisted underwater acoustic communication systems must be designed to not only maximize signal strength in the main lobe but also suppress interference arising from side lobes. The typical 1-bit and 2-bit coding schemes used in RF-RIS may fall short, as the phase resolution of RIS elements is a significant determinant of system performance. According to array processing principles~\cite{van2002optimum}, an RIS with arbitrary reflection coefficients can increase the directional gain of the reflected wave by approximately 23.46\% ($\sim$\,0.9\,dB) over a 2-bit coding RIS and diminish side lobe interference by around 50\% ($\sim$\,3\,dB).

Allowing flexible control over reflection coefficients also enhances the degrees of freedom (DoF) of UA-RIS, enabling advanced array processing methods that significantly improve the performance of underwater acoustic communication networks. For example, the MCDR can be deployed to create nulls in specific directions, thereby preventing eavesdroppers from intercepting information from the reflected signals. Additionally, the recursive least squares (RLS) algorithm can be used to precisely direct reflected beams towards multiple users, allowing several devices at different locations to simultaneously receive data from the same sender.

\subsection{Challenges of Flexible Reflection Control in UA-RIS}
In RF-RIS, the flexible phase shift is achieved by adjusting the reverse bias voltage applied to the varactor's p-n junction~\cite{pei2021ris, abeywickrama2020intelligent}. However, this approach is impractical for UA-RIS due to the acoustic signals operating at frequencies six orders of magnitude lower than RF communications. Specifically, the reflection coefficient, denoted as $\Gamma$, in a transmission line is defined as:
\begin{equation}
\label{eq:0aw2}
  \Gamma = \frac{Z_L-Z_0}{Z_L+Z_0},
\end{equation}
where $Z_L$ represents the load impedance, and $Z_0$ is the characteristic impedance of the transmission line, typically 50\,$\Omega$ in RF systems.

For instance, adjusting the phase of \,$\Gamma$ from $-$45$^\circ$\ to $-$90$^\circ$\ requires changing the load impedance from $-j$121\,$\Omega$ to $-j$50\,$\Omega$. At the 2.4\,GHz RF frequency, the capacitance of the load capacitor needs to vary from 0.55\,pF to\,1.33 pF. This range falls within the capability of many commercial varactor diodes~\cite{nte2012nte, 830ser2007zetex}. In contrast, if the frequency is reduced to 25\,kHz for long-distance acoustic communication, the required capacitance of the load capacitor must change from 52\,nF to 128\,nF to achieve the same phase adjustment. This range significantly exceeds the capabilities of varactor diodes, which typically have capacitance values below 1\,nF.

\subsection{Previous Solution of Flexible Reflection Control and its Limitations}
In our previous work \cite{luo2024experimental}, we introduced a synthetic reflection-based RIS to synthesize acoustic waves with customizable amplitude and phase. This approach utilized two reflectors to generate in-phase and quadrature components. In contrast, the new multilayer structure proposed in this paper requires only one reflector. We will next provide a detailed comparison of the two methods.

\begin{figure}[htb]
\centerline{\includegraphics[width=6.5cm]{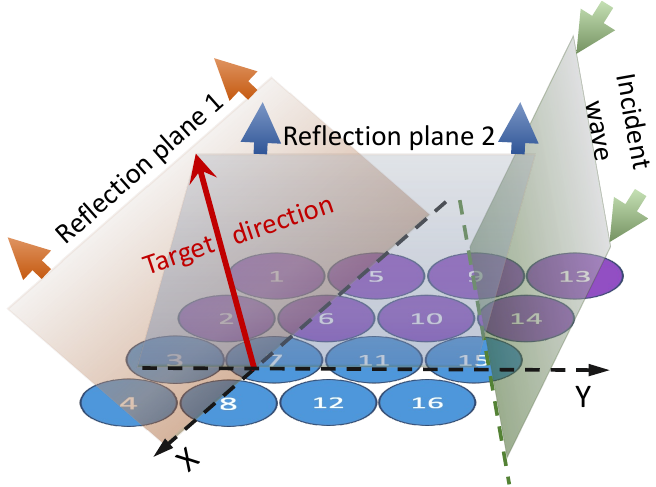}}
  \caption{Synthetic reflection implemented with separated reflectors.}\label{fig:PreWork}
\end{figure}

Fig.~\!\ref{fig:PreWork} illustrates the synthetic reflection scheme from \cite{luo2024experimental}, where two adjacent reflectors sharing the same Y-coordinate are paired. For example, reflector 5 paired with reflector 6, collectively act as a single unit on reflection plane 1. This is possible because their projected positions along the Y-axis are aligned, ensuring that positional differences do not introduce additional phase shifts on that plane. During operation, these paired reflectors produce orthogonal reflection coefficients, allowing the synthesis of reflected waves with arbitrary phases on plane 1 through appropriate adjustments of the amplitude ratio between these orthogonal components.

Using this method, two groups of reflectors indicated by purple and blue in Fig.~\!\ref{fig:PreWork} are formed. Each group consists of four reflector pairs, forming four collective reflectors along the Y-axis. These collective reflectors can generate flexible amplitude and phase of the reflection coefficients. By varying the phase differences among neighboring collective reflectors, the angle of reflection plane 1 can be effectively altered.

In an ideal scenario, the reflected wave would be superposed along a single line oriented toward the target, instead of spreading over a plane. As illustrated in Fig.~\!\ref{fig:PreWork}, this is achieved by configuring two sets of linear arrays along the X and Y axis to form two reflection planes, with their intersection defining the array's directivity.

However, in the UA-RIS proposed in our previous research~\cite{luo2024experimental}, all reflection coefficients are determined solely by the angle of the reflection plane 1 and the incident wave, leaving no additional DoF to establish a second reflection plane, leaving no additional degrees of freedom to form a second reflection plane. This design limitation prevents pairing reflectors along both the X and Y axes, restricting synthetic reflection to one-dimensional gain without enhancing directivity in the orthogonal dimension. For an $N\!\times\! N$ array, this reduces the performance to that of $N/2$ independent linear arrays, each with $N$ acoustic reflectors.

To improve UA-RIS performance, a new structure is necessary to flexibly control the phase and the amplitude of reflected waves. The criteria for this structure are: (a) The design should achieve optimal or near-optimal main lobe gain and side lobe suppression, approaching the performance of an ideal two-dimensional RIS. (b) It should readily adapt to varying environmental conditions, such as variation in intensity and angle of the incident wave, temperature changes, and deployment depths, thereby expanding the application range of UA-RIS. (c) The new structure should be economically feasible, enabling large-scale deployment.



\section{Hardware Architecture}
\label{sec:Hardware}
This section outlines the hardware design of the UA-RIS. The multilayer structure of the acoustic reflector is introduced first. Subsequently, we present the circuit designed to generate in-phase and quadrature-reflected waves for effective synthetic reflection.

\subsection {Multilayered Acoustic Reflector}
\label{sec:MutiRef}
In ML-ARIS, we leverage the acoustic coupling property inherent in mechanical waves to implement synthetic reflection within a single reflector. This property facilitates the transfer of mechanical vibrations across the medium, enabling energy propagation into subsequent layers. Unlike electromagnetic fields, where a conventional RF reflector acts as a conductive boundary that prevents wave penetration beyond the initial layer, the multilayer structure proposed in this work is specific to UA-RIS and fundamentally differs from existing solutions developed for EM-based applications.

\begin{figure}[htb]
\centerline{\includegraphics[width=8.0cm]{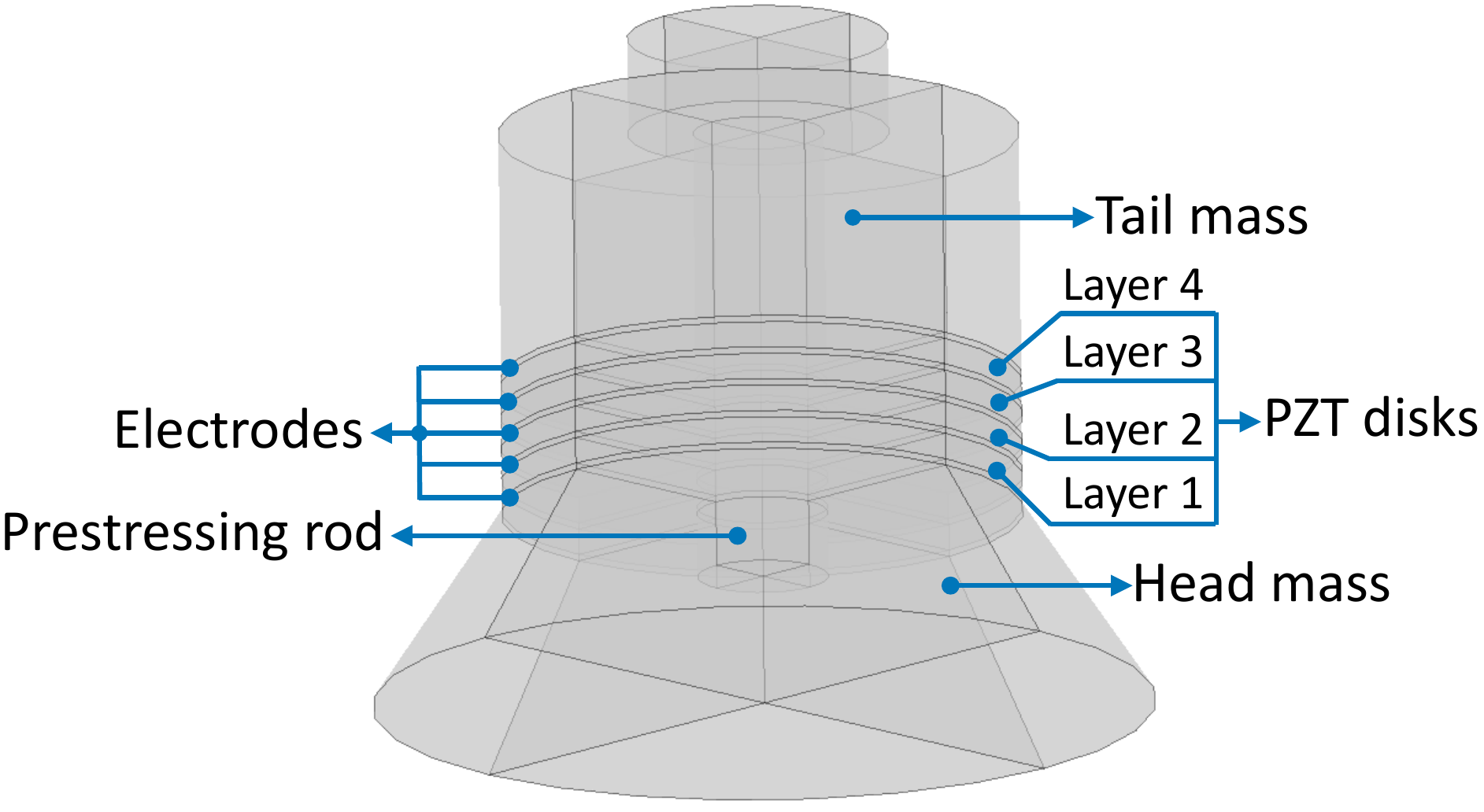}}
  \caption{The structure of a multilayered acoustic reflector.}\label{fig:3DStr}
\end{figure}

The structure of the acoustic reflector designed for the ML-ARIS system is illustrated in Fig.~\!\ref{fig:3DStr}. This reflector employs a Tonpilz-style configuration to achieve efficient electroacoustic conversion in the mid-frequency band~\cite{wilson1985introduction, butler2016transducers}. Within the reflector, a head mass composed of aluminum alloy promotes effective oscillation along the central axis while minimizing stress on other components. A stack of PZT ceramic disks is sandwiched between the head and back masses, serving as the core elements that control the reflected waves. To enhance acoustic coupling efficiency, the PZT stack is held under a constant compressive preload by a prestressing rod. At the rear, a tail mass reflects backward-traveling vibrations, ensuring that the reflected waves propagate forward.

It should be noted that, unlike conventional Tonpilz-type acoustic transducers where piezoelectric disks are connected in parallel, each PZT disk in the ML-ARIS reflector is isolated by insulating materials and operates independently. By adjusting the load impedance of the control circuit associated with each disk, both the phase and amplitude of the reflected wave can be manipulated individually. This design offers a level of control flexibility unattainable with standard transducers.

\subsection {Control Circuit Design}
\label{sec:ConCir}
\begin{figure}[htb]
\centerline{\includegraphics[width=8.5cm]{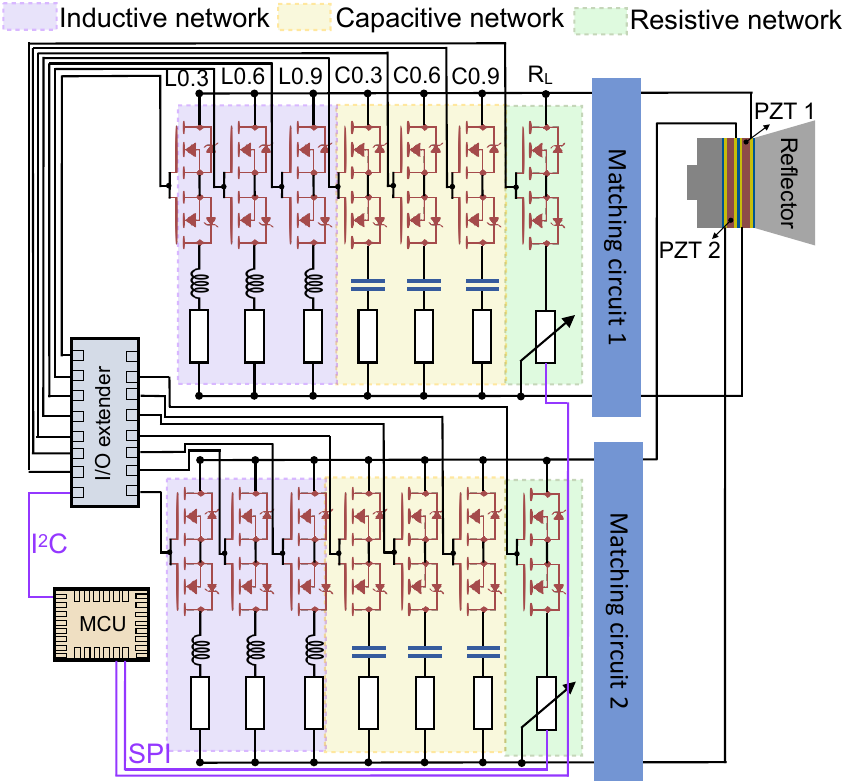}}
  \caption{Control circuit of a two-layer acoustic reflector.}\label{fig:Circuit}
\end{figure}

Fig.~\!\ref{fig:Circuit} illustrates the control circuit developed for a two-layer acoustic reflector. The system employs two identical load networks, each composed of four primary subnetworks: a matching network, a resistive network, a capacitive network, and an inductive network. The TCA9555 I/O extender, interfacing with the ATmega256RFR2 MCU via the I\textsuperscript{2}C protocol, governs the configuration of each load network. This arrangement enables precise manipulation of both the phase and amplitude of the waves reflected by the corresponding PZT disk, thereby facilitating effective synthetic reflection.

The matching network aligns the PZT disk impedance to a specified real value $Z_0$, adapting environmental variations as will be detailed in Section~\!\ref{sec:MatCir}. The resistive network, which includes an AD8403AR50 programmable potentiometer, provides the in-phase component required for synthetic reflection. When the pair of n-type MOS (NMOS) transistors in the resistive network is activated by the I/O extender, the potentiometer is engaged with the matching circuit. Under these conditions, the reflection coefficient of PZT disk 1 is given by
\begin{equation}
\label{eq:masd}
    \Gamma = \frac{R_L-Z_0}{R_L+Z_0},
\end{equation}
where $R_L$ represents the resistance of the programmable potentiometer and $Z_0\!=\!1$\,k$\Omega$ is the characteristic impedance chosen for our circuit design. Here, $Z_0$ is significantly higher than 50\,$\Omega$ used in typical RF systems to accommodate the wiper resistance of the programmable potentiometer. Further details can be found in our previous work \cite{luo2024experimental}.

As indicated by \eqref{eq:masd}, when the load is a resistive component, $\Gamma$ is a real number; therefore, the phases of the reflected and incident waves will be identical or reverse. By adjusting $R_L$ through programming the potentiometer, we can dynamically manipulate the amplitude of the in-phase component.

The capacitive and inductive networks are designed to generate the quadrature component for synthetic reflection. Each subnetwork comprises three stages. Connecting different stages of the subnetwork to the matching circuit yields a reflection coefficient with a consistent phase of $-$90$^\circ$\ (capacitive load) or 90$^\circ$\ (inductive load) and three selectable amplitudes increasing stepwise from 0.3 to 0.9.

\subsection {How Synthetic Reflection Works with ML-ARIS}
\label{sec:ConCir}
Consider an incident wave represented by $\cos(\omega t\!+\!\phi_{in})$, where $\omega$ is the angular frequency, $t$ is time, and $\phi_{in}$ is the initial phase of the incident wave. The desired reflected wave is given by $A_r\cos(\omega t\!+\!\phi_{in}\!+\!\phi_r)$, with $A_r\!\in\! [\,0, 1\,]$ and $\phi_r\!\in\! [-\pi, \pi]$ denoting the amplitude and phase shift of the reflected wave, respectively. Using trigonometric identities, the reflected wave can be expressed as  
\begin{equation}\label{eq:a0qf}
    \begin{aligned}
       & \;A_r\cos(\omega t\!+\!\phi_{in}\!+\!\phi_r) \\
       =& \;A_r\cos\phi_r\cos(\omega t\!+\!\phi_{in})\!- \!A_r\sin\phi_r\sin(\omega t\!+\!\phi_{in}) \\
       =& \;A_r\cos\phi_r\cos(\omega t\!+\!\phi_{in})\!+\! A_r\sin\phi_r\cos(\!\omega t\!+\!\phi_{in}\!\!+\!\!\frac{\pi}{2}).
    \end{aligned}
\end{equation}
As indicated in (\ref{eq:a0qf}), the first term is the in-phase component, maintaining the same phase as the incident wave, while the second term is the quadrature component, which introduces a $-$90$^\circ$\ phase shift relative to the incident wave.

For the two-layer acoustic reflector shown in Fig.~\!\ref{fig:Circuit}, the resistive network of the first PZT disk is activated to produce the in-phase component. From (\ref{eq:masd}), the resistance of the potentiometer is set as  
\begin{equation}
\label{eq:ueie}
  R_L =\left( \displaystyle\frac{1+A_r\cos\phi_r}{1-A_r\cos\phi_r} \right)\!Z_0.
\end{equation}

To generate the quadrature component, the capacitive or inductive network of the second PZT disk is enabled. Specifically, if $\sin\phi_r\!\!\geq\!\!0$, the inductive network is activated; otherwise, the capacitive network is used. Finally, the subnetwork stage that reproduces the amplitude of the reflective wave most close to $A_r\sin\phi_r$ is connected to the matching circuit.

Using the aforementioned method, the PZT disks within the acoustic reflector can produce two orthogonal waves with adjustable amplitudes. These waves are then combined in water, forming a spherical wave with a prescribed phase and magnitude for beam steering.

In a conventional Tonpilz-style acoustic transmitter, multiple PZT disks are connected in parallel and driven by the same electrical signal, resulting in uniform mechanical deformation across all disks. In contrast, for the multilayered reflector proposed here, the mechanical deformation varies among layers because the phases of the electric fields reflected by the load networks of adjacent PZT disks are orthogonal. Consequently, in practical implementations, the reflection signals generated within different layers exhibit a certain degree of coupling rather than being entirely independent. The influence of this coupling on the accuracy of the reflected waves will be examined through simulations and tank experiments, as detailed in Section~\ref{sec:PerEva}.


\section{Enhanced Matching Network}
\label{sec:MatCir}
This section presents an enhanced matching network, designed in response to the varying impedance of acoustic reflectors in the dynamic underwater environment. A simplified circuit model for underwater acoustic reflectors is provided, followed by a detailed discussion on the architecture of the new matching network. We then explore various methods used to optimize its performance across different conditions.

\subsection{Necessity for an Enhanced Matching Network}
\label{sec:WhyMat}
In RF-IRS systems, the impedance of the reflector  changes exclusively with frequency. Under these conditions, a wide-band matching circuit proves highly effective in aligning the impedance of the reflector with that of the load. However, for underwater acoustic reflectors constructed from PZT material, the impedance changes not only with frequency but also in response to water temperature, incident angle, and deployment depth. To ensure efficient operation of the ML-ARIS across various application scenarios, the matching circuit must be wide-band and adaptable to environmental changes.

\hspace{0.2cm}
\subsubsection{Impedance variation with environment and production}
\begin{figure}[htb]
\centerline{\includegraphics[width=7.5cm]{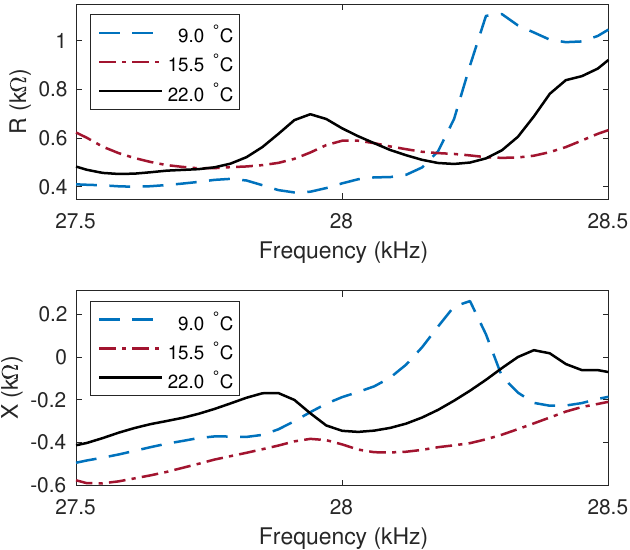}}
  \caption{Impedance variation of PZT disk 1 with frequency and temperature.}\label{fig:impTmp}
\end{figure}

In Fig.~\!\ref{fig:impTmp}, an Agilent E5071C vector network analyzer (VNA) is employed to investigate how the impedance of a PZT disk varies with signal frequency and water temperature. The structure of the reflector used in the test has been introduced in Fig.~\!\ref{fig:3DStr}. The resonant frequencies of all disks are around 28.2\,kHz at 20$^\circ$C. In the figure, the symbols $R$ and $X$ denote the real and imaginary components of the impedance at the first PZT disk, respectively.

As illustrated in Fig.~\!\ref{fig:impTmp}, both $R$ and $X$ exhibit substantial changes at the resonant frequency with varying temperatures. For instance, at a water temperature of 9$^\circ$C and a signal frequency of 28.2\,kHz, $R$ is 678\,$\Omega$ and $X$ is 142\,$\Omega$. As the temperature rises to 22$^\circ$C, $R$ decreases to 494\,$\Omega$, and $X$ shifts to $-$247\,$\Omega$. As a result, with the water temperature rising from 9$^\circ$C to 22$^\circ$C, the magnitude of the PZT disk impedance at 28.2\,kHz increases by 20\%, and the reactance of the impedance transitions from capacitive to inductive.

The impedance of the PZT material also changes with water depth. Specifically, the capacitance of PZT, which depends on their geometry and dielectric properties, is particularly sensitive to these changes. Increased water depth results in heightened pressure, which imposes mechanical stress on the material. This stress compresses the thickness of the material, consequently increasing its capacitance, which in turn leads to a reduction in the imaginary component of the impedance.

In addition to environmental effects (pressure and temperature), manufacturing and assembly tolerances will introduce unit-to-unit and layer-to-layer variability. In ML-ARIS, these production variances affect PZT disk's dielectric/mechanical parameters and resonant frequency, as well as variations in assembly preload and bonding-layer thickness. These factors shift the effective impedance and can therefore alter the achieved reflection coefficient if the load/matching states are selected purely from nominal design values.

 \hspace{0.2cm}
 \subsubsection{Impedance variation with incident angle}
For a typical RF-RIS operating in free space, the direction of incoming waves does not significantly alter the reflector's input impedance. In contrast, for an acoustic reflector, the impedance varies noticeably with the incident angle due to changes in boundary conditions and how sound couples into or out of the piezoelectric material at different angles.

Moreover, the variation in impedance with incident angle may differ across the individual PZT layers. This phenomenon can be validated through a COMSOL multiphysics simulator. Specifically, in the simulation, a reflector comprising four PZT layers is constructed. Each layer measures 6\,cm$\times$0.48\,cm and is separated by an acrylic plastic layer 0.2\,mm in thickness. The incident signal is a monochromatic plane wave with a frequency of 41\,kHz and an intensity of 1\,Pa. 

\begin{figure}[htb]
\centerline{\includegraphics[width=8.5cm]{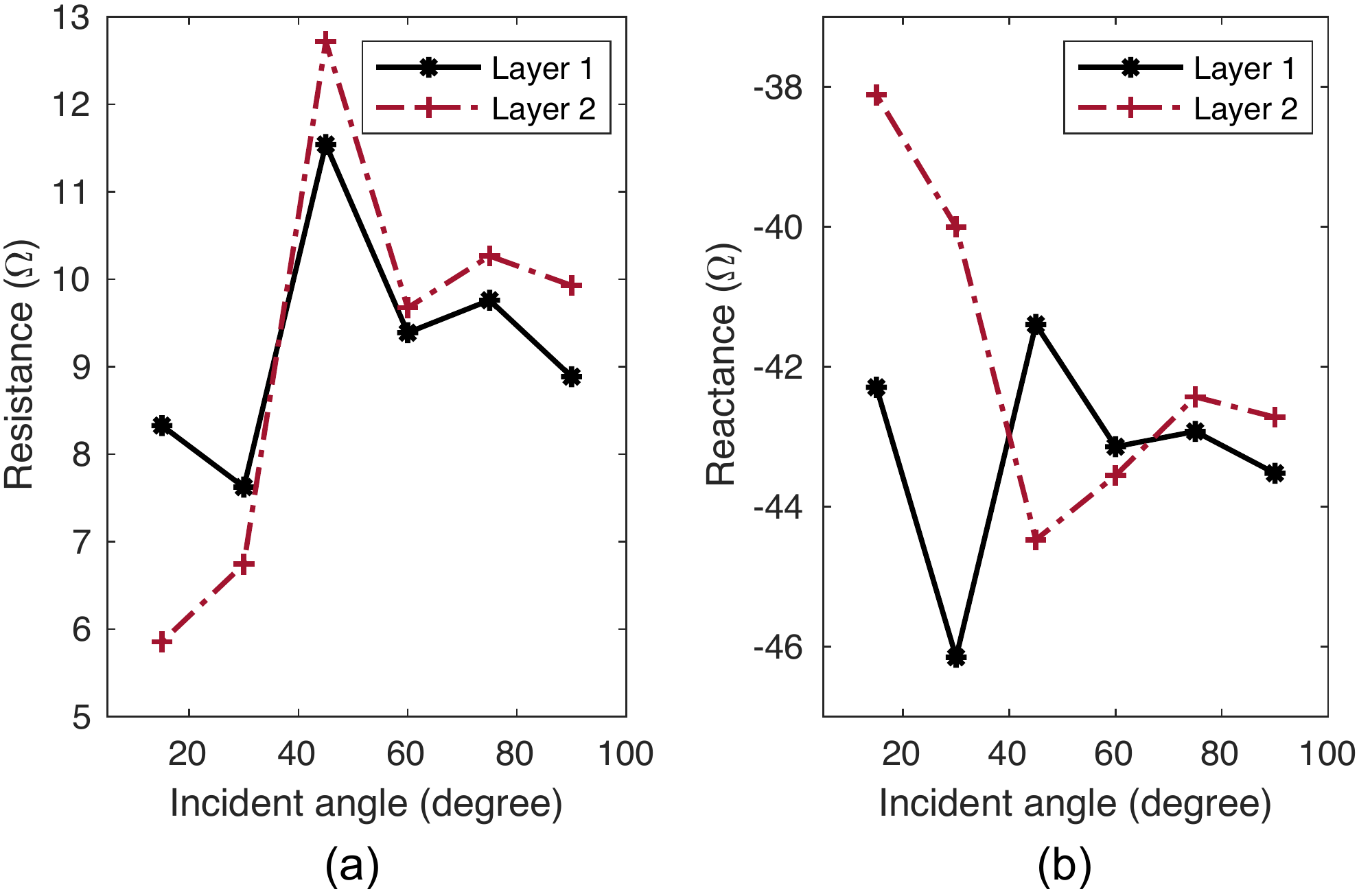}}
  \caption{Impedances of PZT layers under varying incident angles of acoustic waves: (a) Resistance component. (b) Reactance component.}\label{fig:ImAngle}
\end{figure}

In Fig.~\!\ref{fig:ImAngle}, we vary the direction of incoming waves and measure the resistance and reactance for the first two layers of the PZT stack. The impedance in this figure is calculated from the ratio of the open-circuit voltage to the short-circuit current induced by the incoming waves. As illustrated, both the resistance and reactance of the PZT impedance demonstrate non-monotonic variations with changes in the incident angle. For example, at an incident angle of 90$^\circ$, the resistance and reactance for PZT layer 1 are 8.9\,$\Omega$ and $-$43.5\,$\Omega$, respectively. When the incident angle is reduced to 45$^\circ$, the resistance increases to 11.5 $\Omega$, a 29.2\% increase from the initial value. Reducing the angle further to 30$^\circ$, the reactance decreases to $-$46.1\,$\Omega$, a 6\% reduction from the 90$^\circ$\ incident angle.

For PZT layer 2, at an incident angle of 90$^\circ$, the resistance and reactance are 9.9\,$\Omega$ and $-$42.7\,$\Omega$, respectively. Changing the angle to 45$^\circ$\ leads to an increase in resistance to 12.7\,$\Omega$, marking a 28.3\% rise from the 90$^\circ$ situation. When the angle is further decreased to 30$^\circ$, the reactance rises to $-$38.1\,$\Omega$, showing a 9.7\% increase compared to the 90$^\circ$ scenario.

The variation in impedance across different layers arises from distinct acoustic boundary conditions and coupling effects. For instance, the first layer of the PZT stack is in direct contact with water, while the second layer faces the first PZT disk. In this configuration, the refraction, reflection, and scattering of sound waves at the interface between these layers depend on the incident angle. Moreover, in a multilayer stack, the upper layers encounter the incoming wave directly, whereas deeper layers receive a wave that has already been modified by the preceding layers. As a result, the stress distribution among different layers is inconsistent, particularly when the incident wave is not perpendicular to the reflector.

\begin{figure}[htb]
\centerline{\includegraphics[width=9cm]{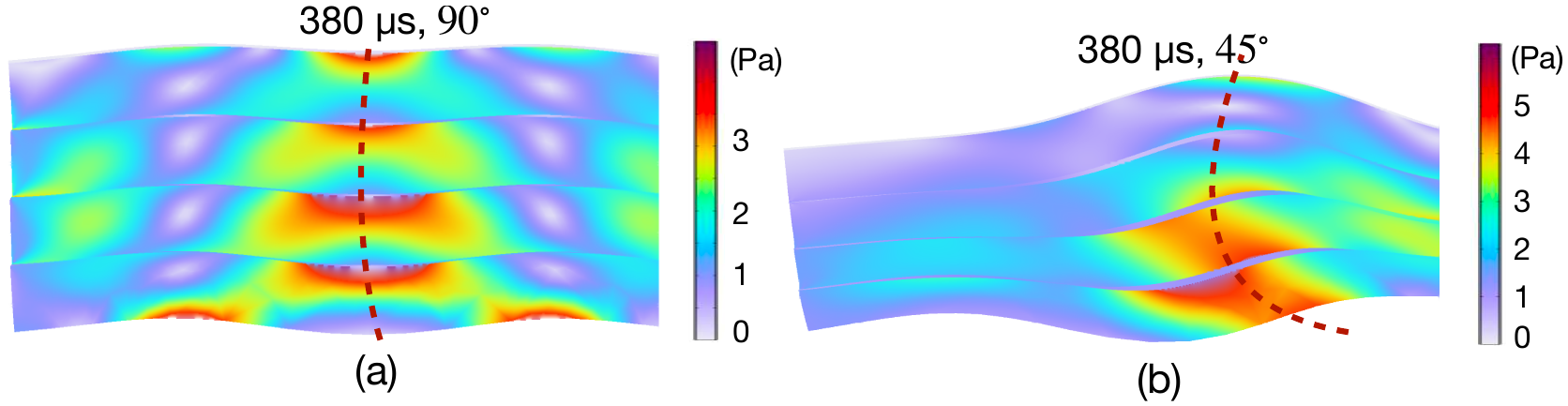}}
  \caption{Stress variation across 4 layers of the PZT stack, measured at 380\,$\mu$s. (a) Incident angle: 90$^\circ$. (b) Incident angle: 45$^\circ$.} \label{fig:StressTime}
\end{figure}

In Fig.~\!\ref{fig:StressTime}, we present the stress distribution across four layers of the PZT stack when stimulated by monochromatic plane waves incident from two different directions at 380\,$\mu$s. The mechanical wave propagation speed in PZT material is approximately 3900\,m/s, and the total thickness of the four-layer stack corresponds to about 0.2 wavelengths. Under these conditions, it can be reasonably assumed that the mechanical wavefront reaches all layers simultaneously. The red dashed line in the figure indicates the maximum instantaneous stress within the stack. As shown in the figure, when the incident angle is 90$^\circ$, the stress distribution across different layers is similar, and the red dashed line appears nearly straight. However, when the incident angle decreases to 45$^\circ$, the stress distribution differs significantly among the layers, causing the red dashed line to form a pronounced arc.

According to the aforementioned analysis, traditional wide-band matching circuits built for RF-RIS are inadequate for handling impedance variations across diverse environmental conditions. This is because their impedance matching depends exclusively on frequency, preventing adaptation to changes in the impedance of an acoustic reflector at any specific frequency. Consequently, for efficient operation in dynamic underwater environments, developing a new matching network is essential for ML-ARIS.

\subsection{Circuit Model of Acoustic Reflector}
\label{sec:RefModel}
\begin{figure}[htb]
\centerline{\includegraphics[width=6cm]{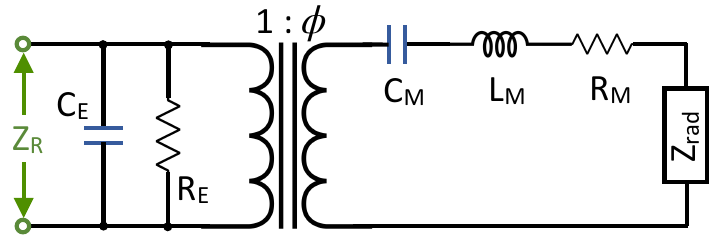}}
  \caption{Electrical equivalent to a PZT disk.}\label{fig:TranModel}
\end{figure}

To design the new matching circuit, it is essential to first develop a circuit model for the PZT disk. Let us denote the impedance of the disk at frequency $f$ as $Z_R(f)$, which can be modeled using electro-acoustic analogies~\cite{lurton2002introduction}. As shown in Fig.~\!\ref{fig:TranModel}, the left part of the model describes the dielectric characteristics of the PZT material, with $R_E$ and $C_E$ indicating the dielectric loss resistance and capacitance of the dielectric plates, respectively. Electrical energy is converted into mechanical energy through a coil transformer with a turns ratio of $\phi$. The right side of the model depicts the mechanical behavior of the reflector, where $R_M$ represents the mechanical loss, $L_m$ and $C_m$ express the dynamic mass and elasticity of the reflector. $Z_{rad}$ is the radiating impedance determined by the geometry of the radiator.

Let $Z_S$ be the equivalent impedance of the secondary winging, where
\begin{equation}
\label{eq:9ia2}
  Z_S=R_M+j(2\pi f L_M-\displaystyle\frac{1}{2\pi fC_M})+Z_{rad}.
\end{equation}
Consequently, the total impedance of the PZT disk can be derived by
\begin{equation}
\label{eq:nmse}
  Z_R = \displaystyle\frac{R_E}{1+\left(2\pi fC_E R_E\right)^2}-j\displaystyle\frac{2\pi fC_E R^2_E}{1+\left(2\pi fC_E R_E\right)^2}+\phi^2 Z_S.
\end{equation}

Near the resonance frequency, the inductive and capacitive reactances of $Z_p$ in (\ref{eq:nmse}) cancel out each other, thus only the resistance of $Z_S$ being reflected onto the primary winding. Additionally,  the dielectric loss resistance ($R_E$) is typically large for underwater acoustic reflectors with high Q-factors, making $2\pi fC_E R_E$ much greater than 1. Under these conditions, the expression for $Z_R$ can be simplified as follows:
\begin{equation}
\label{eq:1k34}
  Z_R \approx \displaystyle\frac{1}{R_E\left(2\pi fC_E \right)^2}+\text{Re}^\prime({Z_S})-j\displaystyle\frac{1}{2\pi fC_E},
\end{equation}
where $\text{Re}^\prime({Z_S})=\phi^2 \text{Re}({Z_S})$.

As analyzed in Section~\!\ref{sec:WhyMat}, the impedance of a PZT disk at any given frequency is not static but varies with environments. Consequently, the resistance and capacitance described in (\ref{eq:1k34}) become variables influenced by factors such as water temperature and pressure. Under specific environmental conditions, a VNA or LCR meter can be used to scan $Z_R$ across various frequencies. Subsequently, a non-linear least squares fitting process can be employed to calculate $R_E$, $C_E$, and $\text{Re}^\prime({Z_S})$. As the environment transitions from condition 1 to condition $m$, a set of $Z_R$ can be recorded.

\subsection{Impedance Estimation}
\label{sec:ImpEes}
As outlined in Section~\!\ref{sec:RefModel}, $\mathbf{Z_R}$ is discrete and can represent the impedance of the PZT reflector in only a limited number of environmental conditions, which is inadequate for designing a matching circuit. Consequently, it is necessary to refine the model of $Z_R$ to approximate the variation in the reflector's impedance across different frequencies and arbitrary environmental conditions.

In $\mathbf{Z_R}$, let $Z^{max}_R(f_r)$ be the element with the maximum magnitude at the resonant frequency $f_r$. The parameters used in (\ref{eq:1k34}) to fit $Z^{max}_R(f)$ are represented by $R^{\alpha}_{E}$, $C^{\alpha}_{E}$, and $\left[\text{Re}^\prime({Z_S})\right]^{\alpha}$. In contrast, the impedance with the maximum magnitude at $f_r$ is represented by $Z^{min}_R(f_r)$; the parameters used to fit $Z^{min}_R(f)$ are denoted as $R^{\beta}_{E}$, $C^{\beta}_{E}$, and $\left[\text{Re}^\prime({Z_S})\right]^{\beta}$, where $R^{\alpha}_{E}\!<\!R^{\beta}_{E}$, $C^{\alpha}_{E}\!<\!C^{\beta}_{E}$, and $\left[\text{Re}^\prime({Z_S})\right]^{\alpha}\!>\!\left[\text{Re}^\prime({Z_S})\right]^{\beta}$.

We now discretize $R_E$, $C_E$, and $\left[\text{Re}^\prime({Z_S})\right]$ into $N_d$ equally spaced values, ranging between their respective minimum and maximum values. The $i$-th value in this series is represented as $R^i_E$, $C^i_E$, and $\left[\text{Re}^\prime({Z_S})\right]^i$, respectively, calculated using the following method:
\begin{equation}\label{eq:y6a2}
    \left\{
    \begin{array}{lll}
        \vspace{0.1cm}
       \!\!\!\! R^i_E =R^{\beta}_{E}-\displaystyle\frac{(i-1)\!\times\!\!\left(R^{\beta}_{E}-R^{\alpha}_{E}\right)}{N_d-1},\\
        \vspace{0.1cm}
       \!\! \!\!C^i_E =C^{\beta}_{E}-\displaystyle\frac{(i-1)\!\times\!\!\left(C^{\beta}_{E}-C^{\alpha}_{E}\right)}{N_d-1},\\
        \!\!\!\!\left[\text{Re}^\prime\!({Z_S})\right]^{\!i} \!\!=\! \left[\text{Re}^\prime\!({Z_S})\right]^{\!\beta} \!\!+\! \displaystyle\frac{\!(i\!-\!1)\!\!\times\!\!\left(\!\left[\text{Re}^\prime\!({Z_S})\right]^\alpha\!\!\!-\!\!\left[\text{Re}^\prime\!({Z_S})\right]^\beta\right)}{N_d-1},\\
    \end{array}
    \right.
\end{equation}
where $i=1,\dots, N_d$.

Based on (\ref{eq:y6a2}) and the simplified circuit model described in (\ref{eq:1k34}), a set of impedances, which is denoted by $\mathbf{Z^{\prime}_R} = \{Z^1_R(f), \cdots, Z^{N_d}_R(f)\}$, can be obtained. We consider $\mathbf{Z^{\prime}_R}$ as an extension of $\mathbf{Z_R}$, representing a preliminary estimation of the reflector's impedance under various environmental conditions. In Section~\!\ref{sec:ArcEnh}, we will explore the development of an enhanced matching network tailored to $\mathbf{Z^{\prime}_R}$. This new network aims to accommodate impedance variations of the PZT disk over a broad range, thus improving the performance of ML-ARIS in dynamic underwater environments.

\subsection{Design of Enhanced Matching Network}
\label{sec:ArcEnh}

\subsubsection{Architecture}
\begin{figure}[htb]
\centerline{\includegraphics[width=9.0cm]{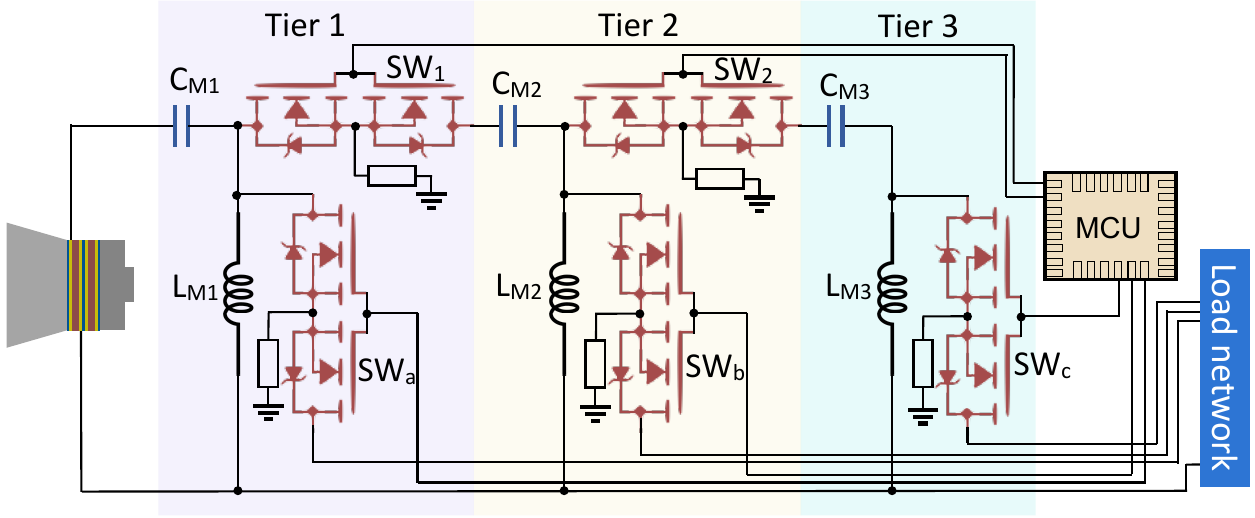}}
  \caption{Architecture of the enhanced matching circuit.}
  \label{fig:MatchingCir}
\end{figure}

In Fig.~\!\ref{fig:MatchingCir}, we illustrate the architecture of the enhanced matching circuit developed for the first layer of the PZT stack in one of the acoustic reflectors. This circuit features a three-tier subnetwork structure, each tier consisting of a high-pass L-type matching circuit. The interconnection between tiers is managed by NMOS transistors $\text{SW}_{1}$ and $\text{SW}_{2}$, which are controlled by the MCU. It is crucial to note that, the operation of the tiers is not independent but cascaded; for instance, activating tier $i$ requires that tiers 1 through $i\!-\!1$ also be activated to achieve the optimal performance. 

In the enhanced matching network, the connection of each tier to the load network is controlled by electronic switches $\text{SW}_{a}$, $\text{SW}_{b}$, and $\text{SW}_{c}$. At any given time, only one switch is active, linking the designated subnetwork to the load. For example, to engage tier-2 to the load, $\text{SW}_{1}$ is first activated to establish a connection through tier-1 to the PZT disk, followed by activating $\text{SW}_{b}$ to directly link tier-2 to the load.

During operation, each tier in the matching network is cascaded with previous tiers to align with the PZT disk impedance within a specified range. To facilitate the automatic switching of the output tier in response to changes in the reflector's impedance, the MCU periodically connects the digital potentiometer (AD8403AR10, as illustrated in Fig.~\!\ref{fig:Circuit}) to the load network and sets its resistance to $Z_0$ specified in Section~\!\ref{sec:ConCir}. Upon detecting the preamble of the acoustic signal, the matching network is sequentially linked into the circuit tier by tier. The closer a tier's output impedance matches $Z_0$, the higher the power delivered from the incident signal to the potentiometer. Therefore, when transitioning between tiers of the matching network, the MCU measures the average voltage at the potentiometer using the internal analog-to-digital converter (ADC). It then selects the tier that generates the maximum voltage on the potentiometer for impedance matching.

Next, we will elaborate on the optimization of parameters in each tier of the matching network, ensuring that the characteristic impedance of the transmission line remains consistently at $Z_0$, unaffected by variations in the impedance within dynamic underwater environments.

\vspace{0.2cm}
\subsubsection{Optimization}
When the load impedance is $Z_0$, assume the desired reflection coefficient of each layer in ML-ARIS remains zero throughout the frequency range from $f_L$ to $f_H$ under different environmental conditions. In tier $i$ of the matching circuit, the capacitance and inductance are designated as $C_{M_i}$ and $L_{M_i}$, respectively. The impedance from the PZT disk to tier $i$ of the matching circuit at frequency $f$ is expressed as $Z_{M_i}(f)$. 

For a 3-tier matching circuit, we set $N_d=9$ in (\ref{eq:1k34}). Initially, tier-1 is optimized to align with the first three impedance values from $Z^1_R(f)$ to $Z^3_R(f)$ in $\mathbf{Z^{\prime}_R}$. Subsequently, tier-2 joints in to optimize for the middle three impedance values in $\mathbf{Z^{\prime}_R}$, specifically $Z^4_R(f)$, $Z^5_R(f)$, and $Z^6_R(f)$. Once the configurations for tier-1 and tier-2 are established, tier-3 collaborates with them to address the final three impedance values in $\mathbf{Z^{\prime}_R}$: $Z^7_R(f)$, $Z^8_R(f)$, and $Z^9_R(f)$.

According to Fig.~\!\ref{fig:MatchingCir}, the output impedance of tier-1 is expressed as follows:
\begin{equation}
\label{eq:a2ry}
  Z^i_{M_1}(f) = \left(Z^i_R(f)-j\displaystyle\frac{1}{2\pi fC_{M_1}}\right) \parallel j2\pi fL_{M_1},
\end{equation}
where the symbol ``$\parallel$'' represents the parallel connection. Then, we can formulate the following optimization problem to determine the inductance ($L_{M_1}$) and capacitance ($C_{M_1}$) in tier-1 of the matching circuit:
\begin{equation}\label{eq:hyte}
\begin{array}{lll}
\hspace{-0.3cm}
    \textbf{\emph{P1}}\;\;
    \vspace{0.1cm}
    \underset{C_{M_1},\, L_{M_1}}{\mathrm{arg\,min}}\,\displaystyle{\sum^3_{i=1}}\,\displaystyle{\sum^{f_H}_{f=f_L}}\,\abs{\displaystyle\frac{Z_0-Z^i_{M_1}(f)}{Z_0+Z^i_{M_1}(f)}}^3, \\
    \begin{array}{lll}
    \vspace{0.05cm}
    \hspace{-0.45cm}
    \textbf{s.t.} & \textbf{\emph{C1:}}\quad C_{M_1} > 0,\\
    \hspace{-0.45cm}
    &\textbf{\emph{C2:}}\quad L_{M_1} > 0.
    \end{array}
    \end{array}
\end{equation}

In the optimization problem $\textbf{P1}$, the primary component of the objective function is the cube of the magnitude of the reflection coefficient at the load impedance $Z_0$. The goal of the optimization is to minimize the sum of these values when the PZT disk impedances are $Z^1_R(f)$, $Z^2_R(f)$, and $Z^3_R(f)$ across the target frequency range. Cubing the cost function can amplify the impact of larger errors more significantly than smaller ones. Consequently, deviations of the reflection coefficient from zero are penalized more severely than they would be under the original cost function, thus focusing on reducing larger discrepancies. 

In our design, the simulated annealing algorithm~\cite{van1987simulated} is employed to address the optimization problem described above. By solving for $C_{M_1}$ and $L_{M_1}$ from (\ref{eq:hyte}), $Z^i_{M_1}(f)$ can be obtained through (\ref{eq:a2ry}). Subsequently, the output impedance for tier-2 is derived using $Z^i_{M_1}(f)$, and an optimization problem analogous to $\textbf{P1}$ is formulated to determine $C_{M_2}$ and $L_{M_2}$. This step aims to minimize the reflection coefficients associated with the impedances $Z^4_R(f)$, $Z^5_R(f)$, and $Z^6_R(f)$. Repeating this methodology, $Z^i_{M_3}(f)$ is calculated based on $Z^i_{M_2}(f)$. From here, $C_{M_3}$ and $L_{M_3}$ are determined to optimize the reflection coefficients for the PZT disk impedances at $Z^7_R(f)$, $Z^8_R(f)$, and $Z^9_R(f)$.

The enhanced matching network, together with the associated optimization method, provides a mechanism for calibration-based compensation to reduce the impact of environmental variation and production-induced variability. Specifically, each PZT unit can undergo a one-time production or installation calibration using a VNA or an impedance analyzer to obtain its impedance map across the intended frequency band and expected environmental range. The resulting lookup table (or fitted model) is stored for subsequent control. During operation, the MCU selects the appropriate tier configuration and load state based on the calibrated map combined with real-time environmental estimates, thereby compensating for both environmental variability and production-induced discrepancies.


\section{Performance Evaluation}
\label{sec:PerEva}
This section validates the design of ML-ARIS through both simulation and experimentation. We assess the performance of the enhanced matching circuit and the synthetic reflection capability of the multilayer structure. Results from tank experiments and COMSOL simulations confirm the operational viability of ML-ARIS, demonstrating its effectiveness in real-world applications.

\subsection {Performance of the Enhanced Matching network}
\label{sec:PerEnh}
\begin{figure}[htb]
\centerline{\includegraphics[width=6.5cm]{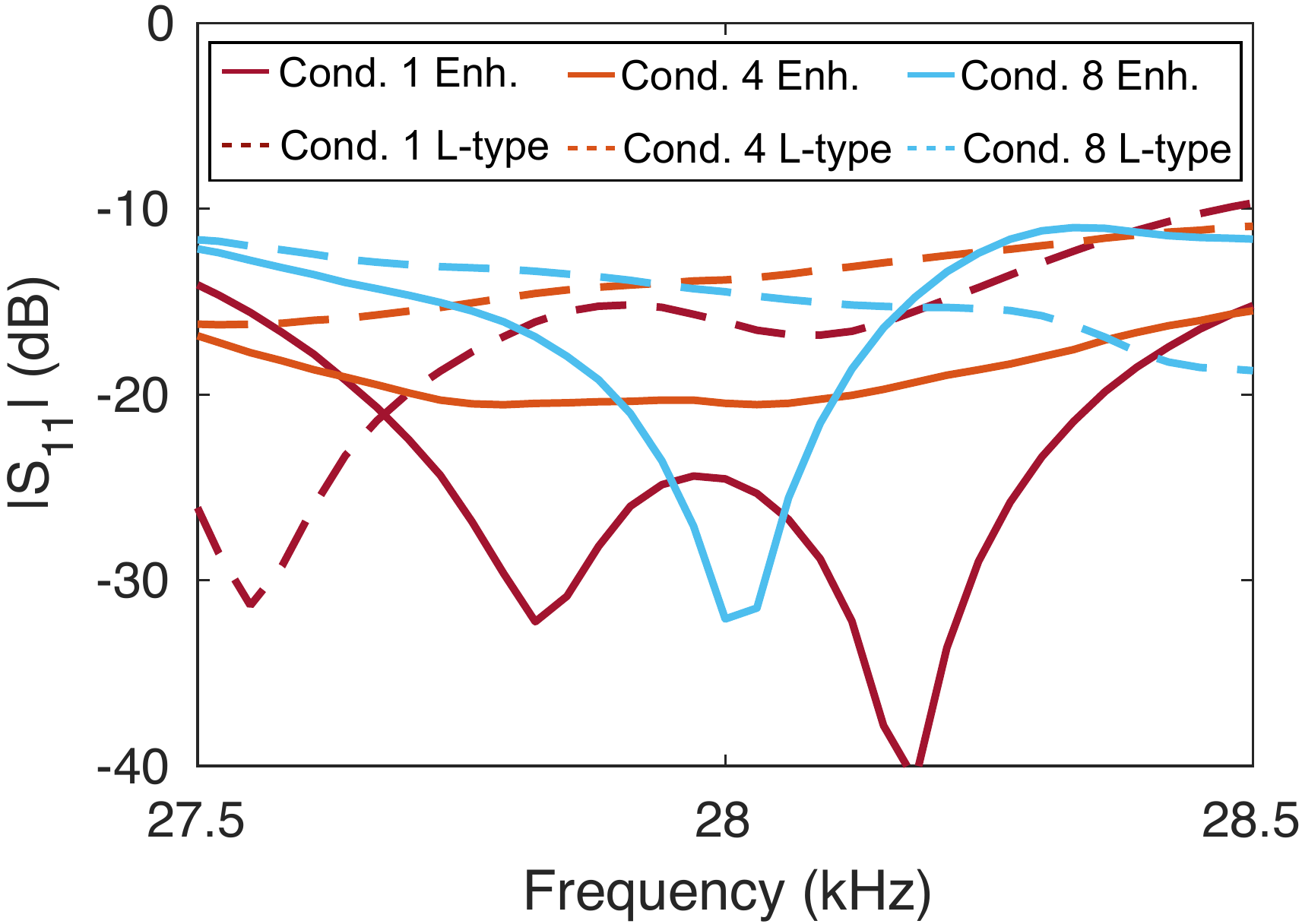}}
  \caption{Comparison of the enhanced matching network and the L-type matching circuit.}\label{fig:TransS11}
\end{figure}

Fig.~\!\ref{fig:TransS11} compares the magnitudes of $\text{S}_{\text{11}}$ between the enhanced matching network and the conventional L-type matching circuit, which are commonly utilized in radio and acoustic communication systems, through experiments. Here, we focus on environmental conditions that have pronounced variations in the impedance of the acoustic reflector. In this comparison, the L-type matching circuit is optimized to align the average resistance and reactance of the reflector at the 28\,kHz resonant frequency with the 1\,k$\Omega$ load.

To determine $\text{S}_{\text{11}}$ for the reflection unit at a 1\,k$\Omega$ target load resistance, the matching circuit is connected to the acoustic reflector. The impedance of the matching circuit is then measured across the selected environmental conditions and frequencies using the Agilent VNA. Subsequently, the resulting magnitude of the $\text{S}_{\text{11}}$ parameter at frequency $f$ is calculated as follows:
\begin{equation}
\label{eq:1j39}
  \abs{\text{S}_{\text{11}}(f)}_i = 20\log\abs{\displaystyle\frac{Z^i_m(f)-1\,\text{k}\Omega}{Z^i_m(f)+1\,\text{k}\Omega}},
\end{equation}
where $Z^i_m(f)$ represents the impedance of the matching circuit under environmental condition $i$. 

As shown in Fig.~\!\ref{fig:TransS11}, within the frequency band of 27.5\,kHz to 28.5\,kHz, the average $\abs{\text{S}_{\text{11}}}$ using the enhanced matching network are consistently lower than those using the L-type circuit across all environmental conditions. While the L-type circuit minimizes $\abs{\text{S}_{\text{11}}}$ at the resonant frequency, it struggles to adapt to variations in the reflector's impedance, resulting in reduced performance in dynamic environments. In contrast, the enhanced matching network employs a cascaded structure that effectively maintains low $\text{S}_{\text{11}}$ magnitudes despite changes in the reflector's impedance. Specifically, under environmental conditions 1, 4, and 8, the average $\abs{\text{S}_{\text{11}}}$ values with the enhanced matching network are $-$21.8\,dB, $-$15.7\,dB, and $-$14.9\,dB, respectively. These values are 3.1\,dB, 2.2\,dB, and 0.5\,dB lower than those achieved with the L-type circuit.

Now, we connect three L-type circuits in parallel to evaluate their performance relative to the enhanced matching network. Each L-type circuit is specifically designed to match impedance within a designated range in this configuration. The final output is dynamically switched among the three circuits based on the reflector's impedance. This contrasts with the enhanced matching network, where three-tier L-type circuits are cascaded, and the output of each tier is dependent on the input from the previous one. In the parallel matching network, the circuits function independently. The design of the parallel matching network is structured around the following optimization problem:
\begin{equation}\label{eq:wpre}
\begin{array}{lll}
\hspace{-0.3cm}
    \textbf{\emph{P2}}\;\;
    \vspace{0.1cm}
    \underset{C_{P_j},\, L_{P_j}}{\mathrm{arg\,min}}\,\displaystyle{\sum^{3j}_{i=3j-2}}\,\displaystyle{\sum^{f_H}_{f=f_L}}\,\abs{\displaystyle\frac{Z_0-Z^i_{P_j}(f)}{Z_0+Z^i_{P_j}(f)}}^3, \\
    \begin{array}{lll}
    \vspace{0.05cm}
    \hspace{-0.45cm}
    \textbf{s.t.} & \textbf{\emph{C1:}}\quad C_{P_j} > 0,\\
    \hspace{-0.45cm}
    &\textbf{\emph{C2:}}\quad L_{P_j} > 0,
    \end{array}
    \end{array}
\end{equation} 
where $j=1,2,3$, representing the $j$-th L-type circuit. In the above optimization problem, $C_{P_j}$ and $L_{P_j}$ are the capacitance and inductance, respectively, of circuit $j$. The target load resistance is $Z_0\!=\!1$\,k$\Omega$. The output impedance of L-type circuit $j$, denoted as $Z^i_{P_j}(f)$, is determined based on the reflector's impedance $Z^i_R$, which is calculated using equations (\ref{eq:1k34}) and (\ref{eq:y6a2}).
 
\begin{figure}[htb]
\centerline{\includegraphics[width=6cm]{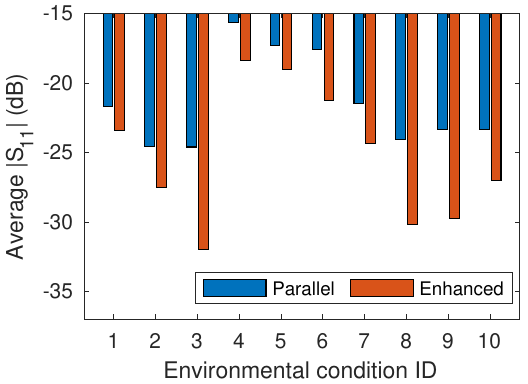}}
  \caption{Average $|\text{S}_\text{11}|$ for enhanced and parallel matching networks under different environmental conditions. The frequency range is from 27.5\,kHz to 28.5\,kHz, and the load impedance is 1\,k$\Omega$.}\label{fig:ParVsCas_new}
\end{figure}

In Fig.~\!\ref{fig:ParVsCas_new}, the resistance and capacitance of the acoustic reflector in various environments range between [200, 450]\,$\Omega$ and [20, 50]\,nF, respectively. The figure illustrates that the performance of the enhanced matching circuit with a cascaded structure is consistently better than that of the parallel configuration under all environmental conditions. Specifically, measurement data indicate that the average $\abs{\text{S}_{\text{11}}}$ values for the enhanced and parallel matching networks are $-$24.1\,dB and $-$20.7\,dB, respectively, with the latter being 3.4\,dB higher than the former.

Furthermore, a comparison between Fig.~\!\ref{fig:ParVsCas_new} and Fig.~\!\ref{fig:TransS11} shows that both the parallel and cascaded matching networks achieve significantly lower $\abs{\text{S}_{\text{11}}}$ values compared to a single L-type circuit. Given that both matching networks require the same number of components (inductors, capacitors, and NMOS switches), yet the cascaded version more effectively matches the reflector's impedance to the target load in dynamic environments,  emerging as the preferred option for constructing our UA-RIS.

\subsection {Verification of Multilayered Structure}
\label{sec:VerIQ}
\subsubsection{Experimental settings}
\begin{figure}[htb]
\centerline{\includegraphics[width=9cm]{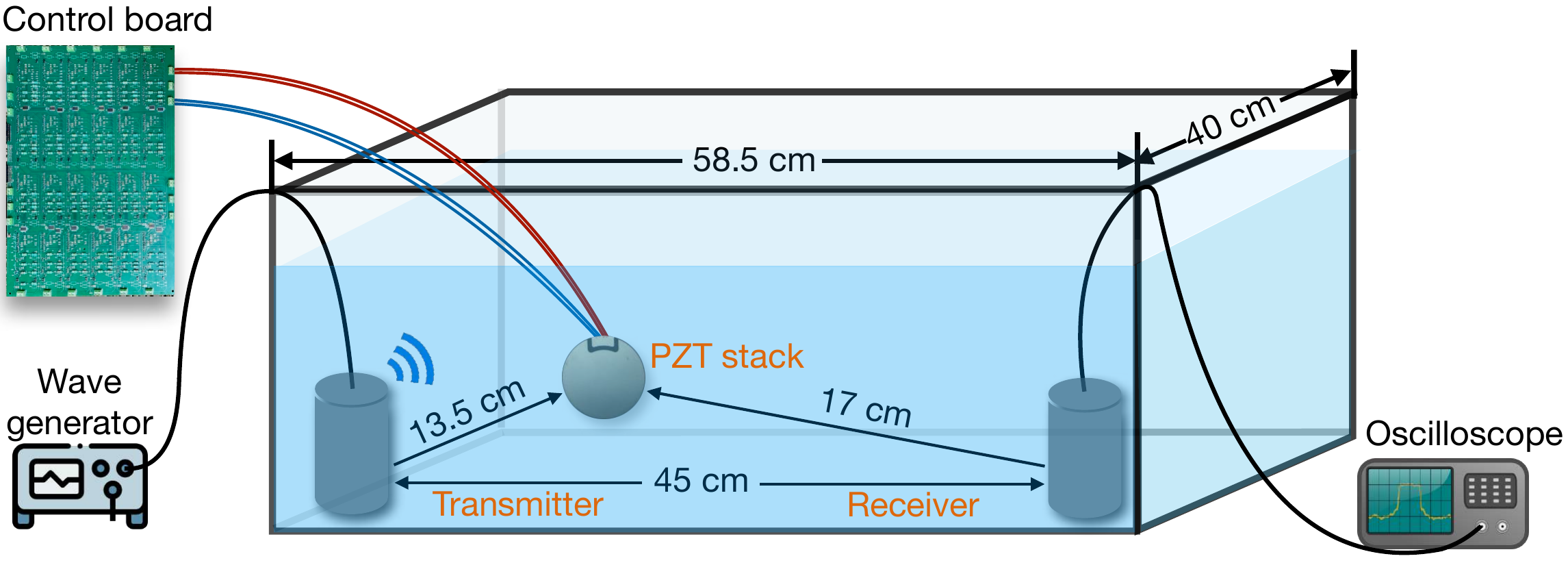}}
  \caption{Settings of tank tests.}\label{fig:TankScen}
\end{figure}

In Fig.~\!\ref{fig:TankScen}, tank experiments are performed to validate the feasibility of synthetic reflection using ML-ARIS, i.e., to demonstrate that the amplitude and phase of the reflector-generated component can be regulated by adjusting the load states of individual layers. In the tests, two BTech Acoustic BT-2RCL~\cite{btech2024btech} omnidirectional acoustic transducers serve as the transmitter and receiver, respectively. The transmitter is driven by a Siglent 1062X arbitrary waveform generator emitting a burst sinusoidal signal consisting of 2,500 cycles. The frequency and peak-to-peak voltage of the source signal are 43.8\,kHz and 20\,V, respectively. The receiver is connected to a PicoScope 4224A oscilloscope for data acquisition.  

The reflectors consist of two PZT-4 disks separated by a polyethylene terephthalate (PET) film. The sound wave propagation speed in the disk is approximately 3900\,m/s; each PZT disk measures \,49.2\,mm diameter\,$\times$\,2.7\,mm thickness and resonates near 44\,kHz. The matching network and load states of the reflectors are controlled by the circuit board shown in Fig.~\!\ref{fig:Circuit}. The conducted experiments focus on the multilayer PZT stack and its load-dependent electromechanical response, rather than a full Tonpilz transducer model. Although Fig.~\!\ref{fig:3DStr} illustrates a Tonpilz-style reflector with head and tail masses for conceptual completeness, the objective here is to isolate the impedance modulation and synthetic reflection behavior enabled by independently loaded PZT layers. For this reason, the head and tail masses are not included in the test as they primarily influence radiation efficiency and resonance characteristics, while the reflection coefficient synthesis demonstrated in this section is governed by the PZT stack and its electrical loading.

\hspace{0.2cm}
\subsubsection{Preprocessing}
In the tank test, the received signals consist of a mixture of three components: (a) the direct wave, (b) waves reflected by the reflector one or more times, and (c) other multipath signals that do not interact with the reflector, where (a) and (c) dominate the received signal. To examine the amplitude and phase generated by the reflected wave, component (b) needs to be separated from the received signal. 

To achieve the above objective, we model two reference signals recorded by the receiver as follows:
\begin{equation}\label{eq:ow46}
    \left\{
    \begin{array}{lll}
        \vspace{0.2cm}
        \!\!r_{opop}(t) = \mathfrak{A}(t)+\!\displaystyle\sum^M_{k=1}\!\left(B_1^{op}e^{j\phi_1^{op}} \!+\!B_2^{op}e^{j\phi_2^{op}} \right)\! s(t\!-\!\tau_k),\\
        \vspace{0.13cm}
        \!\!r_{shsh}(t) = \mathfrak{A}(t)+\!\displaystyle\sum^M_{k=1}\!\left(B_1^{sh}e^{j\phi_1^{sh}} \!\!\!+\!B_2^{sh}e^{j\phi_2^{sh}} \right)\! s(t\!-\!\tau_k),\\
    \end{array}
    \right.
\end{equation}
where $r_{opop}(t)$ and $r_{shsh}(t)$ represent the received signals when the two PZT disks are in open-circuit and short-circuit status, respectively. The term $s(t)$ is the original transmission signal and $\mathfrak{A}(t) \!\!=\!\!\!\sum^N_{i=0}\!A_i e^{j\psi_i}s(t\!-\!\tau_i)$, where $A_i$, $\psi_i$, and $\tau_i$ denote the amplitude, phase, and propagation delay of the channel coefficients associated with the direct wave ($i\!=\!0$) and the $i$th multipath signal ($i\!>\!0$) that do not interact with the reflector. The parameters $B_1^{op}$ and $\phi_1^{op}$ specify the amplitude and phase, respectively, of the wave reflected by the first PZT layer under open-circuit status, while $B_2^{op}$ and $\phi_2^{op}$ represent those for the second PZT layer. 

In \eqref{eq:ow46}, only waves undergoing a single reflection from the reflector are considered, because signals undergoing multiple reflections experience significant attenuation and thus have a negligible impact on the received signal. According to the definition of the reflection coefficient presented in \eqref{eq:0aw2}, under ideal conditions, $B_1^{op}\!=\!B_2^{op}\!=\!1$, $\phi_1^{op}\!=\!\phi^{op}_2\!=\!0$, and  $\phi^{sh}_1\!=\!\phi^{2sh}_2\!=\!\pi$. In this scenario,
\begin{equation}
\label{eq:k4w9}
  \mathfrak{A}(t) = \frac{r_{opop}(t) + r_{shsh}(t)}{2}.
\end{equation}

From \eqref{eq:ow46}, it can be observed that $\mathfrak{A}(t)$ does not depend on the reflector's load status in a stable acoustic environment. Therefore, the waves reflected by the reflector for any load status can be extracted from the received signal using two reference signals. For instance, if the loads of the first and second layers of the PZT stack are $\text{C}_\text{0.6}$ and $\text{R}_{\text{L}}$, respectively, the received signal is denoted by $r_{C_{0.6}R_L}(t)$, while the signal reflected by the PZT stack is given by
\begin{equation}
\label{eq:juyt}
  \mathfrak{B}_{\,C_{0.6}R_L}(t) \!=\!\displaystyle\sum^M_{k=1}\! \left(B^{C_{0.6}}_1e^{j\phi^{C_{0.6}}_1} \!+B^{R_L}_2e^{j\phi^{R_L}_2} \right)\! s(t-\tau_k).
\end{equation}
Based on (\ref{eq:k4w9}), $\mathfrak{B}_{\,C_{0.6}R_L}(t)$ can be calculated as follows: 
\begin{equation}
\label{eq:hvut}
	\begin{array}{lll}
  		\mathfrak{B}_{\,C_{0.6}R_L}(t) \!\!\!\!&=&\!\!\! r_{C_{0.6}R_L}(t) - \mathfrak{A}(t)\\
  		\!\!\!\!&=&\!\!\!r_{C_{0.6}R_L}(t)-\displaystyle\frac{r_{opop}(t)+r_{shsh}(t)}{2}.
  \end{array}
\end{equation}

The expression in \eqref{eq:hvut} is used to extract the signal reflected by the reflector in a multipath environment from experimental measurements. For performance evaluation, the theoretical result of the reflected wave, $\mathfrak{B}^*_{\,C_{0.6}R_L}(t)$, is also needed. According to the circuit design presented in Fig.~\!\ref{fig:Circuit}, under ideal conditions,
\begin{equation}\label{eq:uyt1}
    \left\{
    \begin{array}{lll}
        \vspace{0.2cm}
        \!\!\!B^{C_{0.6}}_1e^{j\phi^{C_{0.6}}_1}\!\!\!\!\!\!\!&=&\!\!\! 0.6\,e^{-j\frac{\pi}{2}},\\
        \vspace{0.13cm}
        \!\!\!B^{R_L}_2e^{j\phi^{R_L}_2}\!\!\!\!\!\!\!\!&=&\!\!\!\displaystyle\frac{R_L-Z_0}{R_L+Z_0},\\
    \end{array}
    \right.
\end{equation}
where $Z_0\!=\!1$\,k$\Omega$ is the characteristic impedance introduced in (\ref{eq:masd}). Substituting (\ref{eq:uyt1}) into (\ref{eq:juyt}), $\mathfrak{B}^*_{\,C_{0.6}R_L}(t)$ is available:
\begin{equation}
\label{eq:uqw3}
  \mathfrak{B}^*_{\,C_{0.6}R_L}(t) \!=\! \left[0.6e^{-j\frac{\pi}{2}}\!+\!\left(\displaystyle\frac{R_L-Z_0}{R_L+Z_0}\right)\right]\displaystyle\sum^M_{k=1}\!s(t\!-\!\tau_k).
\end{equation}

A comparison of $\mathfrak{B}_{\,C_{0.6}R_L}(t)$ with $\mathfrak{B}^*_{\,C_{0.6}R_L}(t)$ enables evaluating the accuracy of the amplitude and phase of the reflected wave. For convenience, we first normalize $\mathfrak{B}_{\,C_{0.6}R_L}(t)$ and $\mathfrak{B}^*_{\,C_{0.6}R_L}(t)$ through $\mathfrak{B}_{\,opop}(t)$. Here, $\mathfrak{B}_{\,opop}(t)$ represents the signal reflected by the PZT stack when both layers are in open-circuit status, i.e., $\mathfrak{B}_{\,opop}(t)\!=\!\sum^M_{k=1}\!\left(B^{op}_1e^{j\phi^{op}_1} \!\!\!+\!B^{op}_2e^{j\phi^{op}_2} \right)\! s(t-\tau_k)$, which can be obtained experimentally as follows:
\begin{equation}
\label{eq:9afy}
  \mathfrak{B}_{\,opop}(t) =\displaystyle\frac{r_{opop}(t)-r_{shsh}(t)}{2} = 2\displaystyle\sum^M_{k=1}\!s(t-\tau_k).
\end{equation}
According to the (\ref{eq:uqw3}) and (\ref{eq:9afy}), the normalized $\mathfrak{B}^*_{\,C_{0.6}R_L}(t)$ can be expressed as follows:
\begin{equation}
\label{eq:iad4}
  \displaystyle\frac{\mathfrak{B}^*_{\,C_{0.6}R_L}(t)}{\mathfrak{B}_{\,opop}(t)} = 0.3\,e^{-j\frac{\pi}{2}}\!+\!\displaystyle\frac{R_L-Z_0}{2\left(R_L+Z_0\right)}
\end{equation}

Finally, when the load status of the two layers are $\text{C}_\text{0.6}$ and $\text{R}_{\text{L}}$, the accuracy of the reflected signal generated by the multilayer reflector in a multipath environment can be estimated by the difference between $\mathfrak{B}^*_{\,C_{0.6}R_L}(t)/\mathfrak{B}_{\,opop}(t)$, calculated by \eqref{eq:iad4}, with $\mathfrak{B}_{\,C_{0.6}R_L}(t)/\mathfrak{B}_{\,opop}(t)$, obtained experimentally via \eqref{eq:hvut}. This methodology can be applied to analyze the amplitude and phase of the reflection signals for arbitrary load status in the multilayered acoustic reflector.

\hspace{0.2cm}
\subsubsection{Experimental results}
\begin{figure}[htb]
\centerline{\includegraphics[width=9.0cm]{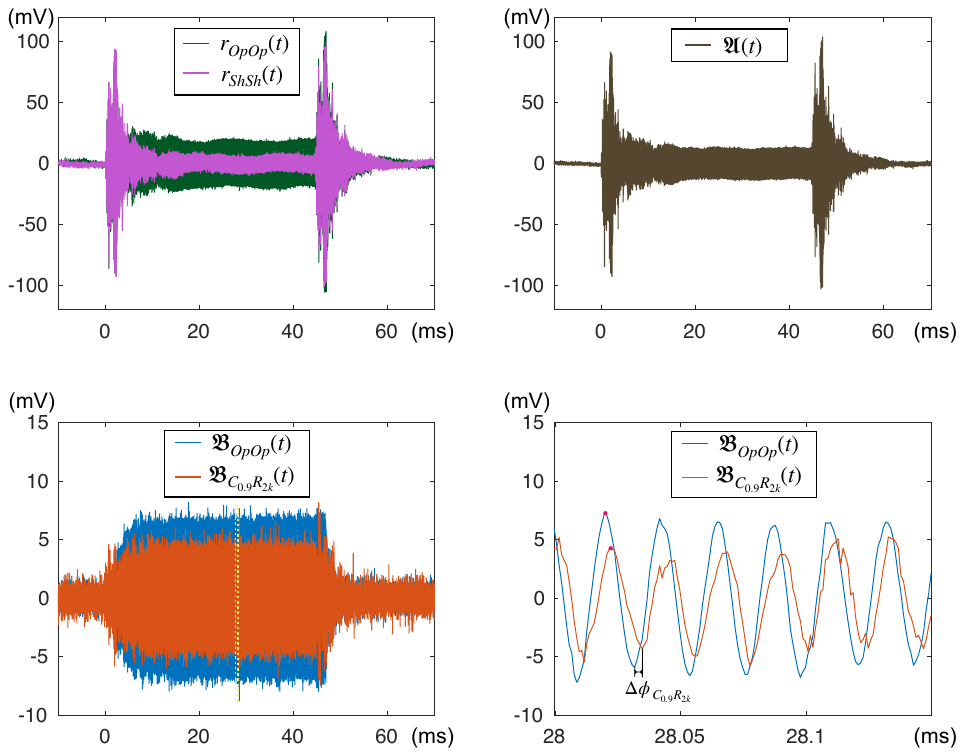}}
  \caption{Waveforms collected in the tank test.}\label{fig:WavForm}
\end{figure}

In Fig.~\!\ref{fig:WavForm}, we present example waveforms collected during the tank experiment. These results are used to calculate the phase and amplitude of the reflected signals when loads of the two layers on the PZT stack are $\text{C}_{0.9}$ and $\text{R}_{2\text{k}}$, respectively.

As shown in Fig.~\!\ref{fig:WavForm}\,(a), the beginnings and ends of the two reference waves exhibit significant fluctuations due to the multipath effect. Once all multipath signals have arrived at the receiver, the combined signal stabilizes between 15 ms and 45 ms. Additionally, it is evident from the figure that changing the load impedance of the reflector significantly affects the strength of the received signal.

Fig.~\!\ref{fig:WavForm}\,(b) shows the waveform of $\mathfrak{B}(t)$, which includes the direct wave and the multiple signals that do not interact with the reflector. As described in (\ref{eq:k4w9}), $\mathfrak{A}(t)$ is computed as half the sum of the two reference waves shown in Fig.~\!\ref{fig:WavForm}\,(a). Fig.~\!\ref{fig:WavForm}\,(c) displays the waves reflected by the PZT stack under two different load conditions, where $\mathfrak{B}_{\,C_{0.9}R_{\text{2k}}}(t)$ and $\mathfrak{B}_{\,opop}(t)$ are obtained from the two reference waveforms through the expressions introduced in (\ref{eq:hvut}) and (\ref{eq:9afy}), respectively. 

To demonstrate the difference between the two reflected signals, Fig.~\!\ref{fig:WavForm}\,(d) provides a partially enlarged view of the area within the yellow dotted frame in Fig.~\!\ref{fig:WavForm}\,(c). In this figure, let $\Delta \phi_{\,C_{0.9}R_{\text{2k}}}$ denote the phase difference between $\mathfrak{B}_{\,opop}(t)$ and $\mathfrak{B}_{\,C_{0.9}R_{\text{2k}}}(t)$. Experimental results indicate that the average $\Delta \phi_{\,C_{0.9}R_{\text{2k}}}$ measured is $-$37.9$^\circ$\ and the amplitude ratio of $\mathfrak{B}_{\,C_{0.9}R_{\text{2k}}}(t)$ and $\mathfrak{B}_{\,opop}(t)$ is 0.51.

\begin{figure}[htb]
\centerline{\includegraphics[width=6.5cm]{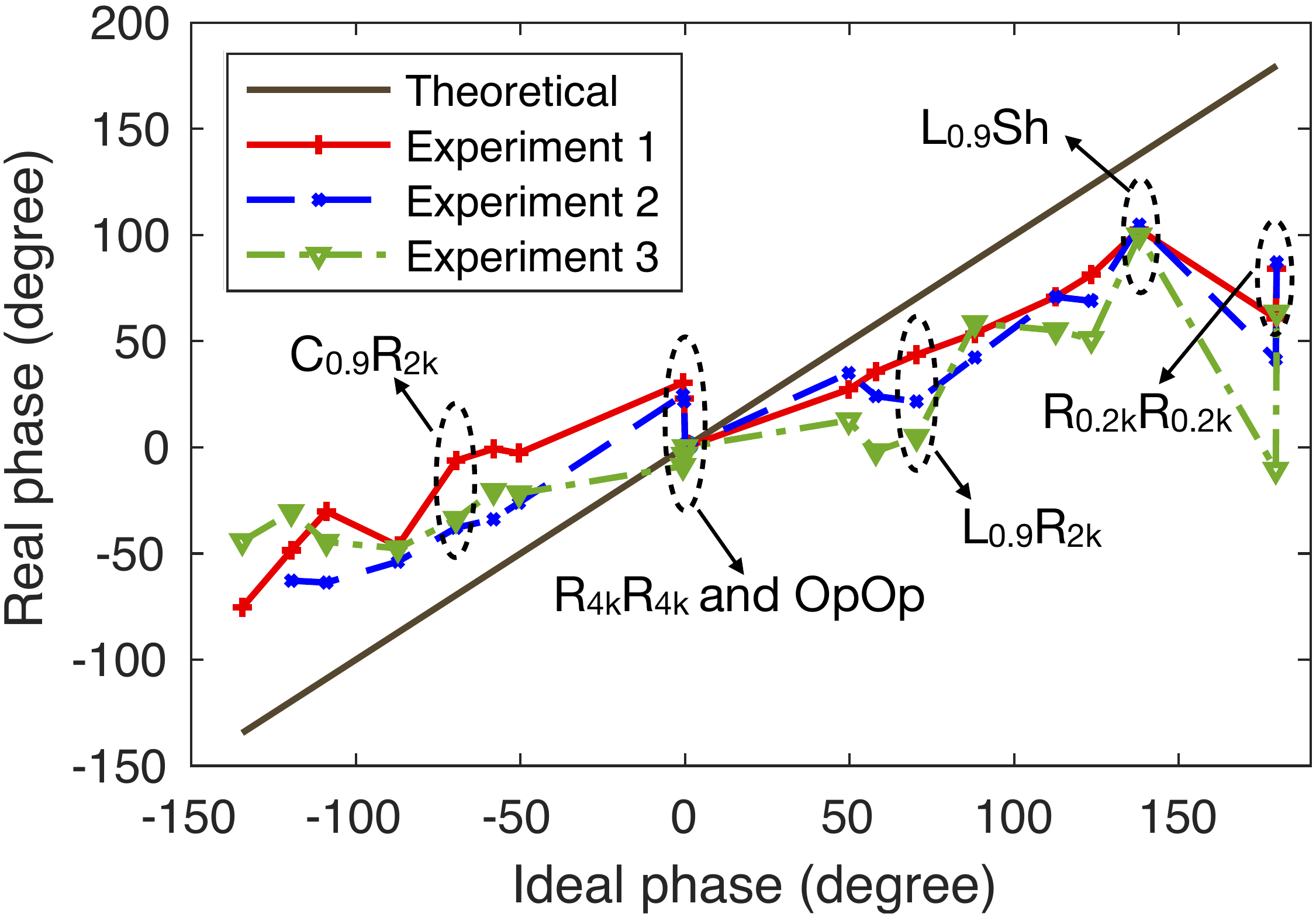}}
  \caption{Phases of reflected waves measured in tank tests.}\label{fig:ExpPhase}
\end{figure}

\begin{figure}[htb]
\centerline{\includegraphics[width=6.5cm]{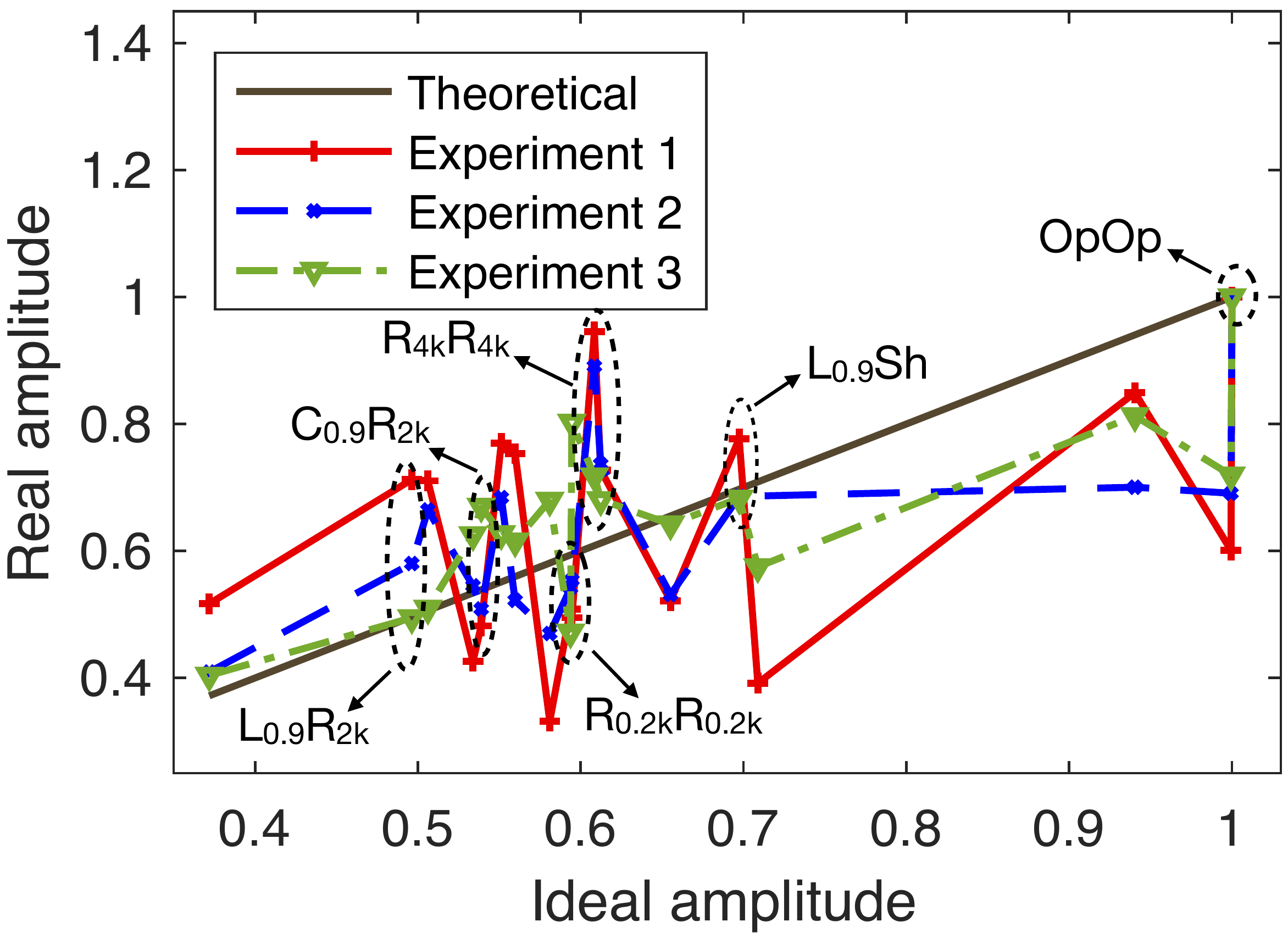}}
  \caption{Normalized amplitudes of reflected waves measured in tank tests.}\label{fig:ExpAmp}
\end{figure}

Figs.~\!\ref{fig:ExpPhase} and \ref{fig:ExpAmp} show the phases and amplitudes of the normalized reflected waves generated by the PZT stack under different load conditions. These figures include results from three tank tests. In each test, the position of the reflector was altered while maintaining other experimental settings constant to evaluate the consistency of the reflected signal. In both figures, the theoretical results are represented by a straight line, computed using the methods described in (\ref{eq:iad4}) with different load impedances at each layer, while the experimental values are obtained via the approach introduced in (\ref{eq:hvut}).

The experimental results demonstrate that the multilayered acoustic reflector successfully generates reflected waves with flexible amplitudes and phases, although the actual results deviate from the theoretical values to some extent. For instance, when loads of the first and second layers of the PZT stack are C0.9 and $\text{R}_\text{L}\!\!=$\,2\,k$\Omega$, respectively (marked as $\text{C}_\text{0.9}$$\text{R}_\text{2k}$ in the figure), the theoretical value of the normalized reflected wave is given by
\begin{equation}
\label{eq:kae6}
  \displaystyle\frac{\mathfrak{B}^*_{\,C_{0.9}R_{2\text{k}}}(t)}{\mathfrak{B}_{\,opop}(t)} = 0.45\,e^{-j\frac{\pi}{2}}\!+\!\displaystyle\frac{2000-1000}{2\left(2000+1000\right)}.
\end{equation}
According to (\ref{eq:kae6}), the phase and amplitude of the normalized $\mathfrak{B}^*_{\,C_{0.9}R_{2\text{k}}}$ are $-$69.7$^\circ$\ and 0.48, respectively.  However, as illustrated in Figs.~\!\ref{fig:ExpPhase} and \ref{fig:ExpAmp}, the average phase of reflected waves measured through three experiments is $-$25.9$^\circ$, and the average amplitude is 0.56.  

Moreover, when the loads of the two layers on the PZT stack are $\text{L}_\text{0.9}$ and short-circuit (marked as $\text{L}_\text{0.9}$$\text{Sh}$ in the figure), the theoretical phase and amplitude of the reflected wave are 138$^\circ$\ and 0.67, respectively. By contrast, the average phase of reflected waves obtained from the tank tests is 102.4$^\circ$, and the average amplitude is 0.71.

From the tank test, it can be observed that both the amplitude and phase of the reflected wave deviate from their theoretical values, primarily due to three factors:  (a) The complex acoustic environment in tank tests, particularly the multipath effect, and limited propagation distance. Multiple reflections by the reflectors cannot be neglected and lead to discrepancies between (\ref{eq:ow46}) and the actual received signal. (b) Theoretical calculations assume a constant characteristic impedance of 1\,k$\Omega$. However, the impedances of the matching circuit and the PZT disks vary with changes in the environment, causing deviations from the theoretical predictions. (c) Mechanical coupling effects, in which the vibration of one PZT disk influences another. This complex interaction was not accounted for in the model, thereby reducing the accuracy of the signal estimation. 

Actually, the coupling requirements in ML-ARIS are inherently contradictory. On one hand, we seek near-perfect coupling between layers so that downstream layers receive as much incident energy as possible and can efficiently transmit their reflected waves through the upstream layers. On the other hand, we require the layers to be mechanically independent so that reflected waves with different even opposing phases do not couple through shared vibrations. Such inter-layer coupling or cross-coupling would distort synthesis.

The experimental results presented in Fig.~\!\ref{fig:ExpPhase} and Fig.~\!\ref{fig:ExpAmp} can be used to estimate the level of cross-coupling. The larger the amplitude and phase deviations from their ideal values, the greater the distortion in the synthesized signal caused by interlayer coupling. For example, when the load of two layers are identical (e.g., $\text{R}_{\text{4k4k}}$ and OpOp), the reflected wave will have the same phase and amplitude. In this case, the measured reflected wave is very close to the ideal case. However, when interlayer vibrations develop phase differences (e.g., $\text{L}_{0.9}\text{R}_{\text{2k}}$ and $\text{C}_{0.9}\text{R}_{\text{2k}}$), mechanical coupling causes the layers to interact, leading the synthesized signal to depart from the ideal response in both amplitude and phase.

Inter-layer coupling in multilayer acoustic reflectors is inherently influenced by a combination of environmental, material, and implementation-dependent factors. In underwater operation, the coupling strength depends not only on the intrinsic mechanical interaction between adjacent PZT layers, but also on external conditions such as water pressure, temperature, surrounding acoustic loading, and even the angle of the incident wave, all of which modify the effective boundary conditions and impedance of the structure. In addition, coupling is frequency-dependent and typically increases near mechanical resonance. Practical realizations further introduce variability through the choice of reflector materials, bonding layers, prestress, and assembly tolerances, as well as manufacturing-induced deviations in PZT properties. As a result, the coupling coefficient should be interpreted as an effective, environment- and configuration-specific metric rather than a universal constant, and its precise value may vary across deployments and operating conditions.

In real implementation, the inter-layer coupling issue can be addressed through calibration. By characterizing the interlayer coupling under representative operating conditions and measuring the resulting amplitude and phase deviations for each load configuration, a compensation map can be built that adjusts the programmed loads to recover the intended in-phase and quadrature components. In effect, calibration shifts the design target from the idealized independent-layer model to the true coupled response, enabling accurate signal synthesis despite residual mechanical interactions.

\subsection {Performance of ML-ARIS}
\label{sec:PerMl}
\subsubsection{Simulation settings}
The simulation is conducted on the COMSOL multiphysics simulator. The reflector comprises 4 PZT-4 layers; only the first two layers are used for synthetic reflection, while the remaining two remain in an open-circuit condition. Each PZT disk measures \,60\,mm diameter\,$\times$\,4\,mm thickness. A monochromatic plane wave with a sound pressure of 1\,Pa at 41.1\,kHz serves as the signal source. Both the reflector and the water medium are surrounded by a perfect matching layer (PML) to mitigate multipath effects.


To enhance the accuracy, the simulation is structured into the following five steps:
\vspace{0.1cm}
\begin{adjustwidth}{-0.71cm}{0cm}
\begin{description}
\setlength{\labelsep}{-0.95em}
\itemsep 0.07cm
\item[a)] Measure the ratio of open-circuit voltage to short-circuit current for each PZT disk under monochromatic plane waves to determine its impedance.
\item[b)] Based on the impedance values obtained in step (a), construct the matching circuit.
\item[c)] Add a load behind the matching circuit and then measure the load voltage under the monochromatic plane wave.
\item[d)] Using the load voltage measured in step (c) and the reflection coefficient of the target load at each layer of the PZT stack, calculate the reflected voltage.
\item[e)] Deactivate the monochromatic plane waves and apply the reflected voltage obtained in step (c) to each PZT layer's matching circuit. Record the resulting pressure waveform, which represents the signal reflected by the reflector under the specified load conditions.
\end{description}
\end{adjustwidth}

\hspace{0.2cm}
\subsubsection{Simulation results of single reflector}
\begin{figure}[htb]
\centerline{\includegraphics[width=6.5cm]{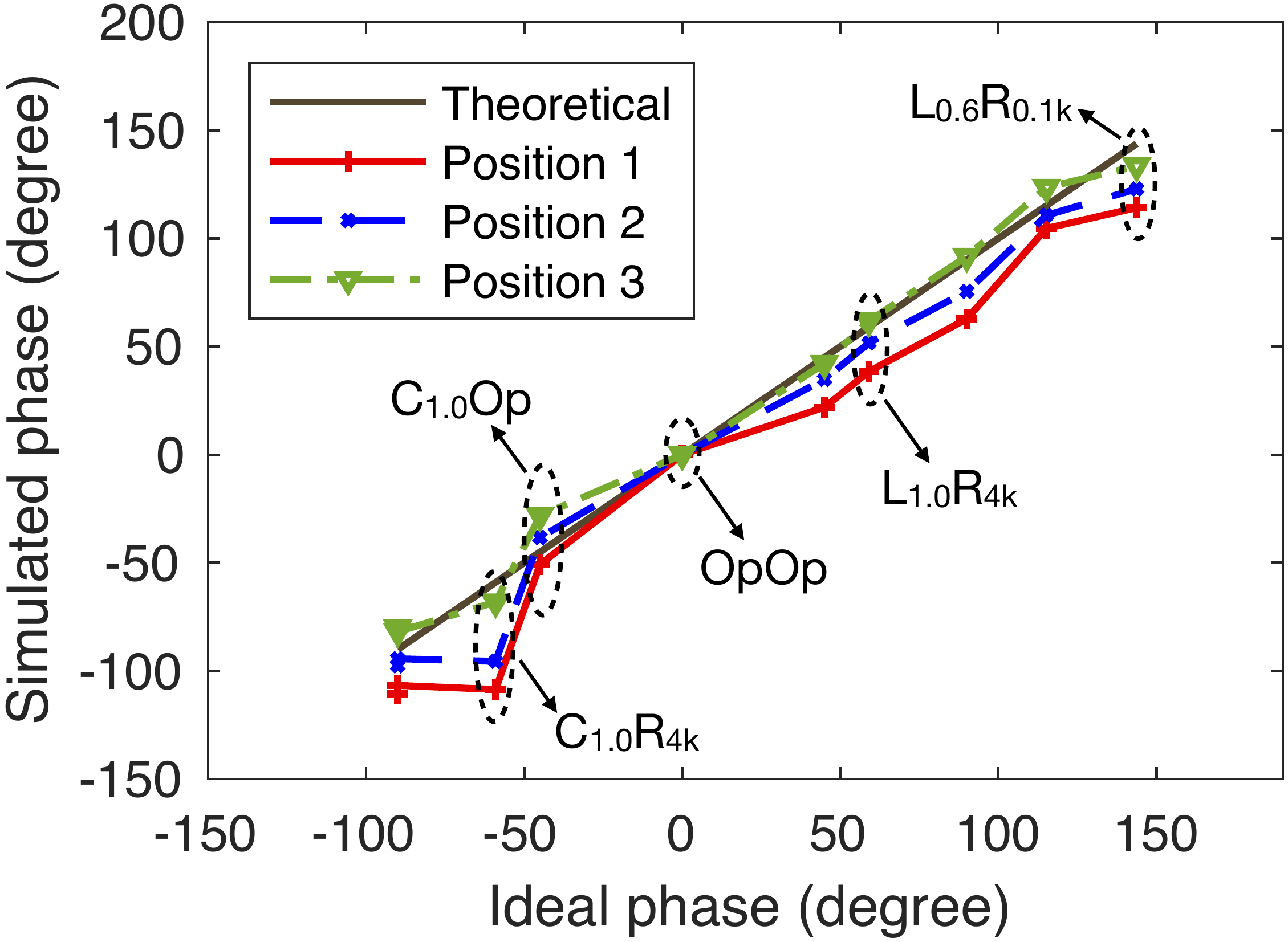}}
  \caption{Phases of reflected waves measured in simulations.}\label{fig:SimPhase}
\end{figure}

\begin{figure}[htb]
\centerline{\includegraphics[width=6.5cm]{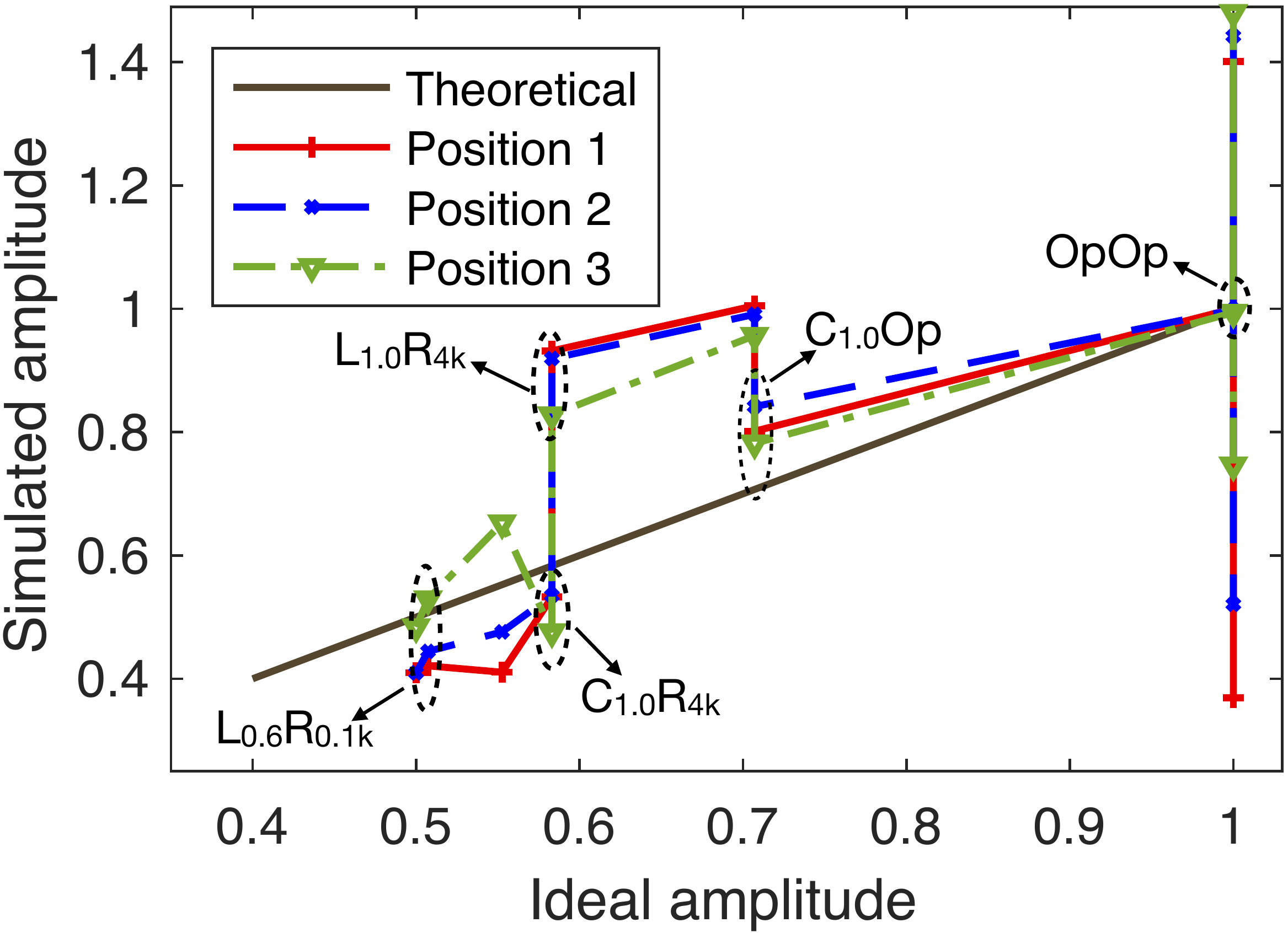}}
  \caption{Normalized amplitudes of reflected waves measured in simulations.}\label{fig:SimAmp}
\end{figure}

Figs.~\!\ref{fig:SimPhase} and \ref{fig:SimAmp} display the phases and amplitudes of the normalized reflected waves obtained from the simulation under various load conditions, respectively. Three measurement probes scattered in the tank are deployed to capture the reflected signal. 

A comparison of Figs.~\!\ref{fig:SimPhase} and \ref{fig:SimAmp} with Figs.~\!\ref{fig:ExpPhase} and \ref{fig:ExpAmp} shows that both the phase and amplitude measured in the simulation correspond more closely with the theoretical results than those from the tank experiments. This improved accuracy is attributed to the use of PML that minimizes multipath effects, thus approximating an anechoic pool environment. Additionally, the five steps in the simulation effectively isolate the incident waves from the reflected waves, eliminating interference from direct waves on the received signal.

\hspace{0.2cm}
\subsubsection{Simulation results of reflector array}
\begin{figure}[htb]
\centerline{\includegraphics[width=9cm]{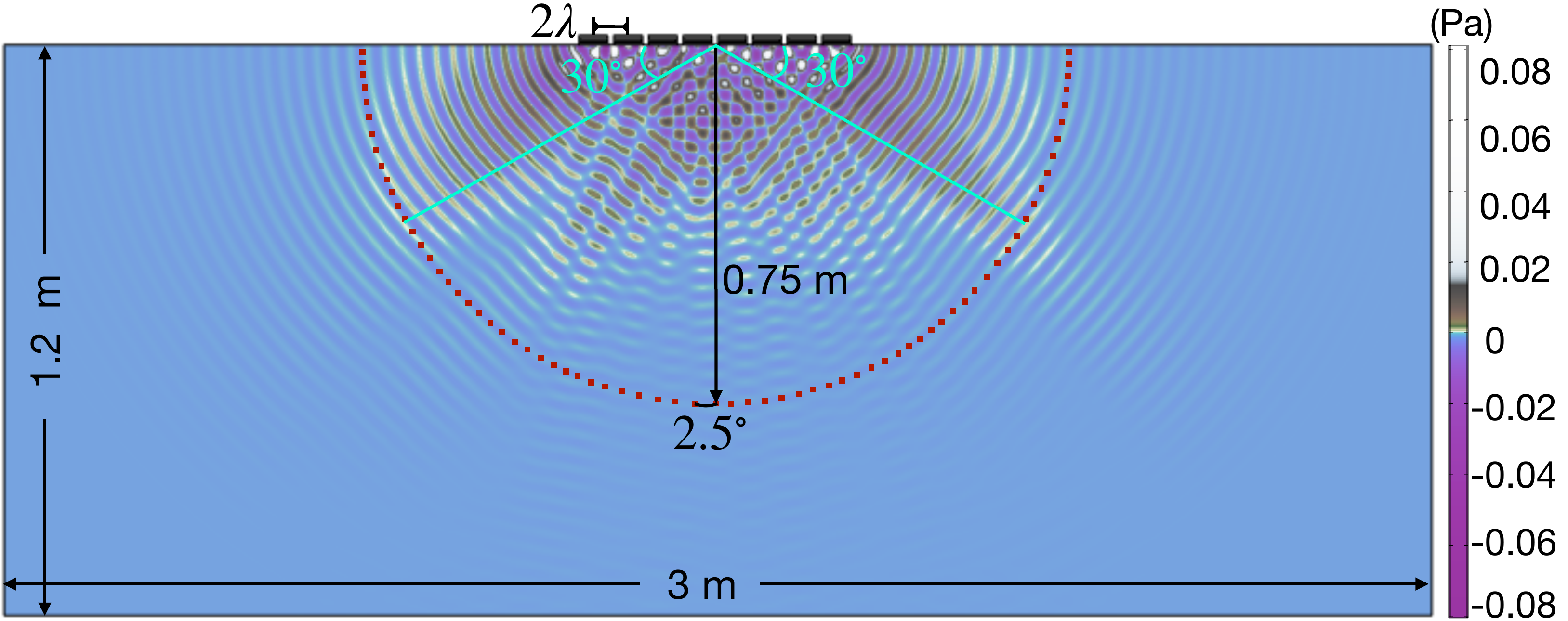}}
  \caption{Reflected beams from an ML-ARIS with 8 reflectors.}\label{fig:SurPres}
\end{figure}

Fig.~\!\ref{fig:SurPres} depicts the reflected beams generated by an ML-ARIS comprising 8 reflectors under a monochromatic plane wave. The simulation configures the spacing between adjacent reflectors to be 2$\lambda$, where $\lambda\!=$3.65\,cm is the wavelength of the plane wave at 41.1\,kHz. This corresponds to an aperture on the order of $(8-1)\times 2\lambda$ along the array axis, i.e., roughly 0.5\,m.  We placed a total of 72 virtual probes, marked as red squares, evenly around a circle to measure the pressure of the reflected wave, with each probe 75\,cm from the center of the ML-ARIS. The incident monochromatic plane wave is eliminated in the figure through the five-step simulation process previously described.

In the simulation, the ML-ARIS is programmed to generate directional gain at 45$^\circ$\ in water using synthetic reflection, with the load network configuration detailed in Table~\!\ref{tab:LdCon}. As shown in the figure, the reflected beam exhibits pronounced directivity near 30$^\circ$, representing a 15$^\circ$\ deviation from the preset value. This discrepancy occurs because the probe lies within the near field,\footnote{$L \!=\! 2D^2 / \lambda$, where $L$ denotes Fraunhofer distance representing the boundary between the near field and far field, and $D$ is the dimension of the ML-ARIS.} preventing the ML-ARIS from being approximated as a point source of negligible volume and the signals generated by each acoustic reflector must be treated as spherical waves rather than plane waves. Furthermore, a grating lobe emerges at approximately 155$^\circ$, resulting from the reflector spacing exceeding one wavelength.


\begin{table}[htp]
\footnotesize
\centering
\setlength{\tabcolsep}{1pt} 
\caption{LOAD CONFIGURATIONS FOR DIFFERENT CODING SCHEMES.}
\label{tab:LdCon}
\begin{tabular}{|
>{\columncolor[HTML]{EFEFEF}}c 
>{\columncolor[HTML]{C0C0C0}}c |
>{\columncolor[HTML]{EFEFEF}}c |
>{\columncolor[HTML]{C0C0C0}}c |
>{\columncolor[HTML]{EFEFEF}}c |
>{\columncolor[HTML]{C0C0C0}}c |
>{\columncolor[HTML]{EFEFEF}}c |
>{\columncolor[HTML]{C0C0C0}}c |
>{\columncolor[HTML]{EFEFEF}}c |
>{\columncolor[HTML]{C0C0C0}}c |}
\hline
\multicolumn{2}{|c|}{\cellcolor[HTML]{D9D9D9} Reflectors}                          & \cellcolor[HTML]{D9D9D9}1 & \cellcolor[HTML]{D9D9D9}2 & \cellcolor[HTML]{D9D9D9}3 & \cellcolor[HTML]{D9D9D9}4 & \cellcolor[HTML]{D9D9D9}5 & \cellcolor[HTML]{D9D9D9}6 & \cellcolor[HTML]{D9D9D9}7 & \cellcolor[HTML]{D9D9D9}8 \\ \hline
\multicolumn{1}{|c|}{\cellcolor[HTML]{EFEFEF}}                     & Layer\,1 & Open                         & $\text{R}_{0.8\text{k}\Omega}$                & $\text{R}_{2.8\text{k}\Omega}$               & $\text{R}_{1.1\text{k}\Omega}$                       & $\text{R}_{0.28\text{k}\Omega}$                & $\text{R}_{19.4\Omega}$               & Short                         & $\text{R}_{9.4\text{k}\Omega}$                       \\ \cline{2-10} 
\multicolumn{1}{|c|}{\multirow{-2}{*}{\cellcolor[HTML]{EFEFEF}IQ}} & Layer\,2 & $\text{R}_{1\text{k}\Omega}$\!\textsuperscript{*}                         & $\text{L}_\text{0.6}$\textsuperscript{\dag}                         & $\text{C}_\text{1.0}$\textsuperscript{\ddag}                        & $\text{L}_\text{1.0}$                        & $\text{C}_\text{1.0}$                        & $\text{L}_\text{0.3}$                        & $\text{R}_{1\text{k}\Omega}$                         & $\text{C}_\text{0.6}$                        \\ \hline
\multicolumn{1}{|c|}{\cellcolor[HTML]{EFEFEF}1-bit}                & Layer\,1 & Short                         & Short                         & Open                         & Short                         & Open                         & Short                         & Short                         & Open                         \\ \hline
\multicolumn{1}{|c|}{\cellcolor[HTML]{EFEFEF}2-bit}                & Layer\,1 & Open                         & Short                         & $\text{C}_\text{1.0}$                        & $\text{L}_\text{1.0}$                        & $\text{C}_\text{1.0}$                        & Open                         & Short                         & $\text{C}_\text{1.0}$                        \\ \hline
\end{tabular}

\vspace{0.2cm}
\textsuperscript{*} Resistive load of 1\,k$\Omega$. 

\textsuperscript{\dag} Inductive load for a reflected-signal phase of $+$90$^\circ$ and magnitude of 0.6,  computed from (\ref{eq:masd}).

\textsuperscript{\ddag} Capacitive load  for a reflected-signal phase of $-$90$^\circ$ and magnitude of  1.0, computed from (\ref{eq:masd}).
\end{table}

It is important to note that the absolute pressure of the reflected wave, as shown in Fig.~\!\ref{fig:SurPres}, is low because only a small portion of the acoustic energy penetrates the PZT, primarily due to the acoustic impedance mismatch between the water and the reflector's head mass. In practical applications, this challenge can be mitigated by incorporating multiple matching layers between the head mass and the water, similar to those employed in medical \cite{dias1996method} or sonar imaging \cite{bian2021ultra}, to significantly reduce the reflection at the bare boundary.

\begin{figure}[htb]
\centerline{\includegraphics[width=6cm]{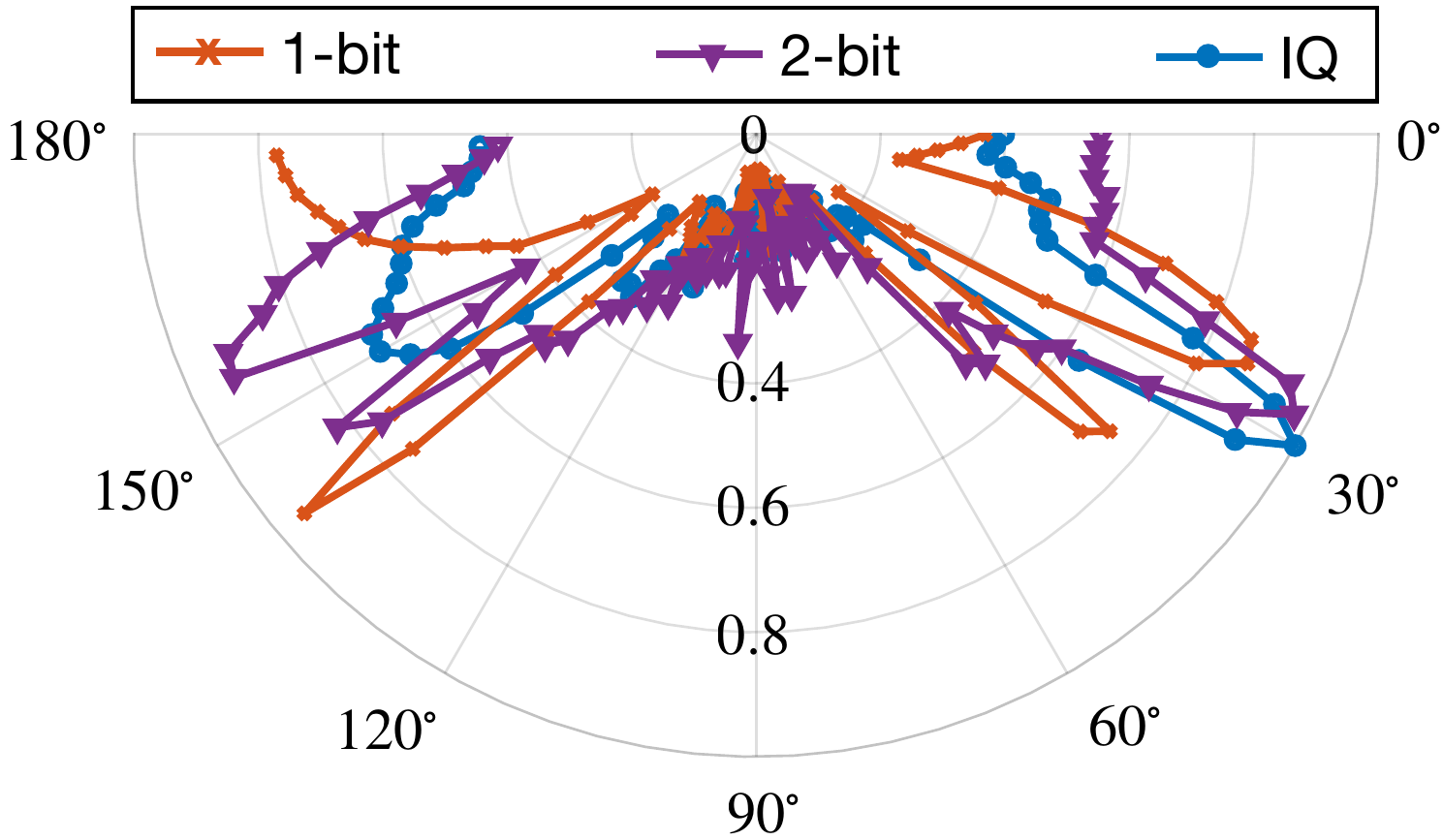}}
  \caption{Comparison of reflected beam generated by different coding methods.}\label{fig:BeamCom}
\end{figure}

In Fig.~\!\ref{fig:BeamCom}, we compare the reflected beams generated by the ML-ARIS under various coding schemes. The configuration is identical to that in Fig.~\!\ref{fig:SurPres}, and the beam amplitude is normalized to the maximum pressure collected by the probes. Each coding scheme is designed to produce a main lobe at 45$^\circ$, with the corresponding load configurations provided in Table~\!\ref{tab:LdCon}. For the 1-bit and 2-bit coding schemes, orthogonal synthesis of the reflected signals is not required; hence, only the first layer of the PZT stack is connected to the load network, while the remaining layers are left open-circuited.

From Fig.~\!\ref{fig:BeamCom}, it can be observed that due to near-field effects, the main lobes of all three beams are deflected to 30$^\circ$, yielding a 15$^\circ$ deviation from the theoretical expectation. Moreover, the side lobe of  reflected beam with synthetic reflection is substantially lower than that of the 1-bit and 2-bit coding schemes. For example, the normalized amplitude of the first side lobe under synthetic reflection is only 0.26 at 47.5$^\circ$, compared with 0.52 at 45$^\circ$\ for 2-bit coding and 0.74 at 40$^\circ$\ for 1-bit coding. Furthermore, the sound pressure of the main lobe under synthetic reflection remains strong, with normalized amplitudes of 1 compared to 0.97, and 0.86 for 1-bit, and 2-bit coding schemes, respectively.

Typical long-range UA-RIS deployments are expected to operate in the far field at the frequencies considered. Accordingly, links with ranges well beyond the Fraunhofer distance fall primarily in the far-field regime, whereas near-field calibration is most critical for close-in configurations (e.g., tank experiments or short-range installations) where near-field beam deviations become non-negligible.

The SNR gains achievable with ML-ARIS are expected to exceed those reported in our prior experimental UA-RIS study \cite{luo2024experimental}. Both ML-ARIS and \cite{luo2024experimental} synthesize reflections using two orthogonal (I/Q) components; however, as discussed in Section.~\!\ref{sec:Moti}, the multilayer architecture effectively doubles the number of controllable reflecting layers per unit, yielding a larger effective aperture (or, equivalently, more degrees of freedom) for the same number of reflection units. For reference, in \cite{luo2024experimental} lake trials, deploying a UA-RIS with 24 reflecting units near the transmitter extended the communication range by up to 28\% and 46\% in deep- and shallow-water environments, respectively. At a fixed transmitter-receiver separation, the UA-RIS increased the received SNR by 2.13\,dB on average, with peak gains up to 2.92\,dB in certain cases. When the UA-RIS was positioned near the receiver, the communication range increased by 40.6\% and 66\% in deep and shallow water, respectively; for a fixed separation, the received SNR improved by 2.56\,dB on average and by as much as 4.2\,dB under specific conditions.


\section{Conclusion and Future Work}
\label{sec:Con}
In this paper, a multilayered reconfigurable intelligent surfaces is proposed for underwater acoustic networks. The stacked architecture enables each individual reflector to generate a reflected wave with flexible amplitude and phase via synthetic reflection. Compared with the multi-bit coding RIS commonly employed in terrestrial communications, the proposed ML-ARIS achieves higher-precision beam steering while minimizing interference to the surrounding environment. To ensure efficient operation in complex ocean settings, an enhanced matching network is also developed to maintain a stable output impedance under varying water temperatures and RIS deployment depths.

The performance of the enhanced matching circuit and the multilayered acoustic reflector is evaluated through both tank experiments and COMSOL simulations around 28\,kHz and 41\,kHz. The results verify the effectiveness of the ML-ARIS in manipulating the phase and amplitude of the reflected wave to support advanced array processing algorithms. This capability establishes ML-ARIS as an enabling hardware platform for future studies of RIS-assisted underwater acoustic communication systems.

UA-RIS is still an emerging research area, and many of the supporting elements required for fielded RIS-assisted links remain open. While this work focuses on the proposed UA-RIS architecture and its hardware foundations, enabling high-resolution reflection control and improving matching robustness under underwater impedance variability, we emphasize that realizing end-to-end communication gains in practical waveguide environments will require further system-level development. In particular, future work should target supporting protocols and signal processing, including channel sounding and tracking, pilot and feedback design, configuration update strategies for time-varying environments and mobility, and practical coordination mechanisms between transmitter, UA-RIS, and receiver to enable reliable RIS-assisted underwater acoustic communications in operational deployments.

\section*{Acknowledgement}

This work is supported in part by the US National Science Foundation under Awards CNS-2048188 and CNS-2016726 and by the Alabama Water Institute Equipment Fund, The University of Alabama.

\bibliographystyle{IEEEtran}
\bibliography{StackUARIS}

\input{bio.tex}

\end{document}

%% file: bio.tex
\begin{IEEEbiography}[\vspace{-0.2cm}{\includegraphics[width=26mm,clip,keepaspectratio]{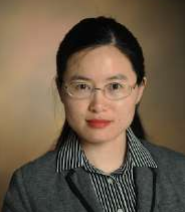}}]
{Dr. Lina Pu} received the B.S. degree in electrical engineering from the Northwestern Polytechnical University, Xi'an, China in 2009 and the Ph.D. degree in Computer Science and Engineering from University of Connecticut, Storrs. Dr. Pu is currently an Assistant Professor at University of Alabama. Her research interests lie in the area of edge computing, RF energy harvesting wireless networks, security in the sustainable IoT, and underwater acoustic networks. She owned IFIP Networking 2013 best paper award.
\end{IEEEbiography}

\begin{IEEEbiography}[{\includegraphics[width=26mm,clip,keepaspectratio]{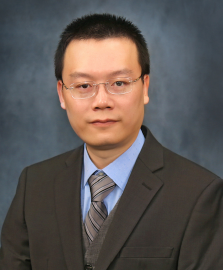}}]
{Dr. Yu Luo} received the B.S. degree and the M.S. degree in electrical engineering from the Northwestern Polytechnical University, China, in 2009 and 2012, respectively. In 2015, he received the Ph.D. degree in computer science and engineering from University of Connecticut, Storrs. Dr. Luo is currently an Associate Professor at Mississippi State University. His major research focus on the IoT, Edge Computing, RF energy harvesting hardware, security in RF energy harvesting wireless networks, and underwater wireless networks. He is a Co-recipient of the Best Paper Award in IFIP Networking 2013 and Chinacom 2016.
\end{IEEEbiography}

\begin{IEEEbiography}[\vspace{0.0cm}{\includegraphics[width=26mm,clip,keepaspectratio]{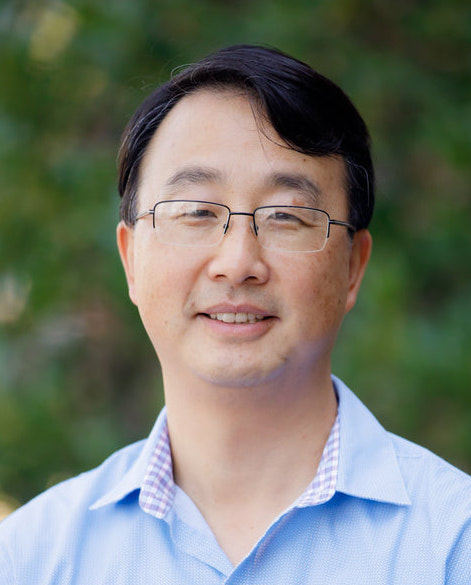}}]
{Dr. Aijun Song} (Senior Member, IEEE) received the Ph.D. degree in electrical engineering from the University of Delaware, Newark, DE, USA, in 2005. From 2005 to 2008, he was a Postdoctoral Research Associate with the College of Earth, Ocean, and Environment, University of Delaware. During this period, he was also an Office of Naval Research (ONR) Postdoctoral Fellow, supported by the Special Research Award in the ONR Ocean Acoustics program. From 2008 to 2015, he was a Research Professor with the University of Delaware. He is
currently an Associate Professor with the Department of Electrical and Computer Engineering, University of Alabama, Tuscaloosa, AL, USA. His research interests include underwater wireless communications and networking, underwater robotics, community-shared open infrastructure for underwater applications, and integrated communications, navigation, and sensing. Dr. Song is a recipient of the National Science Foundation CAREER Award in 2021. He is an Associate Editor for Journal of Acoustical Society of America.
\end{IEEEbiography}